\documentclass[pra,twocolumn,showpacs,amsfonts,amsmath,floatfix]{revtex4-1}

\usepackage{graphicx}
\usepackage[caption=false]{subfig}
\graphicspath{{./images/}}
\usepackage{dcolumn}
\usepackage{bm}%

\usepackage{notoccite}

\input{Qcircuit}

\newcommand{\Bra}[2][]{\left<#2\right|_{#1}}
\newcommand{\Ket}[2][]{\left|#2\right>_{\hspace{-0.1em}#1}}

\newcommand{\ketbra}[3][]{\mathinner{\lvert#2\rangle\langle #3\rvert}_{#1}}
\newcommand{\Ketbra}[3][]{\left|#2\middle>\middle<#3\right|_{#1}}
\newcommand{\proj}[2][]{\ketbra[#1]{#2}{#2}}
\newcommand{\Proj}[2][]{\Ketbra[#1]{#2}{#2}}

\newcommand{\be}{\begin{equation}}
\newcommand{\ee}{\end{equation}}
\newcommand{\ba}[1]{\begin{array}{#1}}
\newcommand{\ea}{\end{array}}
\newcommand{\bal}{\begin{align}}
\newcommand{\eal}{\end{align}}
\newcommand{\baln}{\begin{align*}}
\newcommand{\ealn}{\end{align*}}
\newcommand{\lr}[1]{\left( #1 \right)}
\newcommand{\ddd}{\mathrm{d}}

\begin{document}

\preprint{APS/123-QED}

\title{Polynomial approximation of non-Gaussian unitaries by counting one photon at a time }

\author{Francesco Arzani$^1$}
\email{francesco.arzani@lkb.upmc.fr}
\author{Nicolas Treps$^1$}
\author{Giulia Ferrini$^2$}

\affiliation{$^1$Laboratoire Kastler Brossel, UPMC-Sorbonne Universit\'es, CNRS, ENS-PSL Research University, Coll\`ege de France; CC74, 4 Place Jussieu, 75252 Paris, France; \\
$^2$Institute of Physics, Johannes-Gutenberg Universitaet Mainz, Staudingerweg 7, 55128 Mainz, Germany
}

\date{\today}

\begin{abstract}
In quantum computation with continous-variable systems, quantum advantage can only be achieved if some non-Gaussian resource is available. Yet, non-Gaussian unitary evolutions and measurements suited for computation are challenging to realize in the lab. We propose and analyze two methods to apply a polynomial approximation of any unitary operator diagonal in the amplitude quadrature representation, including non-Gaussian operators, to an unknown input state. Our protocols use as a primary non-Gaussian resource  a single-photon counter.
We use the fidelity of the transformation with the target one on Fock and coherent states to assess the quality of the approximate gate.

\end{abstract}

\maketitle

\section{Introduction}

The use of systems described by an infinite-dimensional Hilbert space, also called continuous-variable (CV) systems, was first considered in the context of quantum computation (QC) in a seminal paper by Lloyd and Braunstein~\cite{lloyd}. This approach is now attracting considerable interest because it allows to generate large resource states deterministically~\cite{yokoyama2013ultra,Yoshikawa2016}. A universal CV quantum computer is defined as a device able to approximate with arbitrary precision any unitary evolution of $m$ modes generated by Hamiltonians that are polynomials of the quadrature operators~\cite{lloyd}. This can be achieved combining single mode Gaussian operations, a single mode non-Gaussian operation and a two-mode entangling gate. Gaussian unitaries are generated by Hamiltonians that are polynomials up to second order in the quadratures and conserve the Gaussian character of the Wigner function. Non-Gaussian evolutions are generated by higher order Hamiltonians. The ability to repeat at will a single non-Gaussian unitary is sufficient to promote the Gaussian set to a universal one~\cite{lloyd}.

In the optical setting, Gaussian unitaries correspond to propagation through passive interferometers, displacements and squeezing operations, which are fairly available in the lab~\cite{suCluster}. Entangling gates can be constructed as multimode Gaussian operations, namely combining squeezers and passive interferometers, such as networks of beam splitters~\cite{irreducible, linClust}. Non-Gaussian gates are the most challenging to implement. Former experimental proposals for their implementation~\cite{oscill} require resources currently out of technological reach~\cite{ghose2007non}. Yet, it has been proven that the use of at least one non-Gaussian operation is necessary in order to build algorithms that could not be efficiently simulated on a classical computer~\cite{eisert, rahimi2015efficient}.

An approach that attracted some attention is based on the fact that a unitary operator can be approximated by the first terms of its Taylor expansion~\cite{marek,yukawa,park,marshall2015repeat}. This is a polynomial in the quadratures of the field, and even though it is not a unitary operator, it can approximate the evolution due to a polynomial Hamiltonian if the evolution time is small enough. 

In this work, we propose and analyze two new methods to implement polynomial gates using squeezed states and detectors that allow to project on a single-photon state, which we will refer to as single-photon counters (SPC). They are inspired by the CV formulation of the measurement-based paradigm for quantum computation (MBQC). In this paradigm, an entangled resource state, called cluster state, is prepared in the beginning and the computation is then driven by local measurements~\cite{raussendorf2003measurement,univCluster, gu2009quantum}. 

Our first approach uses a single photon detector~\footnote{This is ideally a SPC but standard avalanche photodiodes are usually employed in the experiments.} to herald the subtraction of a photon from a beam in a squeezed state, generating an ancillary non-Gaussian state; the building block of the protocol is then completed entangling this state with the input mode and then performing a homodyne detection on the latter. 
Similar methods for engineering non-Gaussian states were already studied, based on the use of ancillary single-photon states and homodyne~\cite{etesse2015experimental,Etesse:14} or heterodyne~\cite{park} detection respectively. 
In the second method that we propose, the input state is coupled to a squeezed ancilla and a single photon is detected in one mode by means of a SPC. As we will see, the two protocols result in different performances, and their applicability therefore will strongly depend on the practical goal, as well as on the specific experimental implementation.
Our schemes may be used either to directly apply a target gate to an unknown input state or to prepare a resource state starting from a known input. 

The work is structured as follows. In Sec.~\ref{sec:poly} we explain the general method to construct a polynomial approximation of a unitary operation. 
Then, after recalling some definitions in Sec.~\ref{sec:BG}, in Sec.~\ref{sec:photosub} we illustrate the first protocol, in which a prototypical circuit for CV-MBQC is fed with a photon subtracted squeezed state instead of a squeezed state, as it would be the case in standard cluster-state computation. We derive the expression of the resulting effective transformation and assess the quality of the gate for a target unitary in terms of fidelity of the transformation on coherent and Fock input states with up to ten photons. In Sec.~\ref{sec:click} the same is done for the second protocol, in which the homodyne detector in the basic circuit for CV-MBQC is replaced by a SPC. Concluding remarks in Sec.~\ref{sec:conclusion} complete the paper.

\section{Polynomial approximation of unitary transformations \label{sec:poly}}

Consider a Hamiltonian operator \be \hat{H} = \mathcal{P} \left(\hat{q}\right)  \ee with $\mathcal{P}$ a polynomial of degree $d$ and $\hat{q}$ the position quadrature. The evolution after a time $t$ under this Hamiltonian is given by the unitary operator \be \label{eq:ev} \hat{U}\lr{t} = \exp\lr{-it\hat{H}}.\ee 
If $t$ is sufficiently small, $ U\lr{t}$ can be approximated by the first terms  of its Taylor expansion in the time parameter 
\be \hat{U}\lr{t} \simeq \hat{U} ^{(n)}\lr{t} = \sum_{j=0}^n \frac{\lr{-it\hat{H}}^j}{j!}\ee 
which is itself a polynomial in $\hat{q}$ of degree $l = d\times n$ and can be decomposed in a product of monomials in the $\hat{q}$ quadrature 
\be\label{eq:development} \hat{U} ^{(n)} \lr{t}=  \prod_{j = 1} ^l \lr{\hat{q} - \lambda_j(t)}, \ee 
where each $\lambda_j$ is a complex number~\cite{park}. 

We want to provide protocols, requiring currently available technology, that allow achieving evolutions of the form of Eq.~(\ref{eq:development}), thereby approximating arbitrary polynomial evolutions (\ref{eq:ev}).
The building block of our protocols will be the non-unitary transformation 
\be \label{eq:genTeff} \hat{T}_\mathrm{eff} = \mathcal{A}\lr{\hat{q}}\lr{\hat{q}-\lambda} \ee 
where $\mathcal{A}\lr{\hat{q}}$ has the form $\exp\lr{-a\lr{\hat{q} - b}^2}$, $\hat{q}$ is the amplitude quadrature of the field, $a$ and $b$  are real numbers. 
The value of $\lambda$ at each realization of the circuit will have to match the $\lambda_j(t)$ in Eq.~(\ref{eq:development}). 
As we will see, in an experimental scenario $\lambda$ depends on tunable parameters, and in one protocol on the output of a homodyne measurement. The factor $\mathcal{A}\lr{\hat{q}}$ is an undesired attenuation of the wave function that determines the range of values of $q$ for which the protocols reproduce a polynomial. This range tends to the whole real axis in the limit of infinite squeezing resources.

By comparing Eqs.~(\ref{eq:development}) and~(\ref{eq:genTeff}) we see that a polynomial approximation of degree $l$ requires applying the effective transformation Eq.~(\ref{eq:genTeff}) $l$ times. We will show that this can be obtained by using either $l$ photon-subtracted squeezed states and $l$ homodyne detections, or $l$ single photon detections, depending on the method used. 

Chaining the effective transformation to achieve $\hat{U}^{(n)}\lr{t}$ comes at the expense of applying the product of the attenuations $\mathcal{A}_j \lr{\hat{q}}$ at each step, where a subscript $j$ has been added, because the parameters $a_j$ and $b_j$ characterizing the attenuation depend in general on the step, as well as on the experimental conditions and the target unitary.  As a consequence, the resulting transformation \be \hat{ \mathcal{T} } = \prod_{j=1} ^l \hat{T}_\mathrm{eff} \lr{j} \ee can be divided in two parts, one being the product of Gaussian envelopes on the position wave function of the input and the second consisting of a polynomial approximating $ \hat{U} ^{(n)} \lr{t} $. 

This transformation is not unitary and obviously differs from the target transformation, so one needs a criterion to evaluate how good the approximation is. A possible choice is to use as a figure of merit the fidelity between the output state obtained with the effective transformation and the result that one would obtain applying the target gate. We consider the case in which the input is a pure state $\Ket{\psi}$. The ideal unitary target gate then produces a pure state $\hat{U}\Ket{\psi}$. In general, however, the output state of our approximated gate will be a mixed state, which we denote here by $\rho$. The fidelity is then~\cite{NChuang} \be \label{eq:fidelityGen} \mathcal{F}= \sqrt{\Bra{\psi}\hat{U}^\dagger  \rho \hat{U} \Ket{\psi} }.\ee If both output states are pure, this reduces to the overlap \be \label{eq:fidelityPure} \mathcal{F} = \left|\Bra{\psi}\hat{U} ^\dagger \hat{\mathcal{T}}\ket{\psi} \right|.\ee The fidelity will generally depend on $\psi$. To test the performance of our protocols we will compute $\mathcal{F}$ on input Fock states and coherent states. 

As it is a widely studied non-Gaussian operation, that allows promoting the Clifford set to a universal set of gates for CV-QC~\cite{gu2009quantum}, we will take the so called cubic phase gate \begin{equation}
\hat{\gamma}\lr{\nu} = \exp\lr{i\nu\hat{q}^3}
\end{equation}  as the target gate. We will compare it to its third order expansion in $\nu$ \be
\label{eq:cubic} 
e^{i\nu\hat{q}^3} \approx \mathbb{I}+i\nu\hat{q}^3 = \lr{q-\lambda_1}\lr{q-\lambda_2} \lr{q-\lambda_3}, 
\ee as obtained chaining three effective transformations of the form in Eq.~(\ref{eq:genTeff}). The roots of the polynomial are $\lambda_1=-i/\nu^{-\frac{1}{3}}$, $\lambda_2=-\lr{-1}^{-\frac{1}{6}}/\nu^{-\frac{1}{3}}$, $\lambda_2=-\lr{-1}^{-\frac{5}{6}}/\nu^{-\frac{1}{3}}$. 

\begin{figure}
    \subfloat[]{
        \centering
        \includegraphics[width=0.24\textwidth]{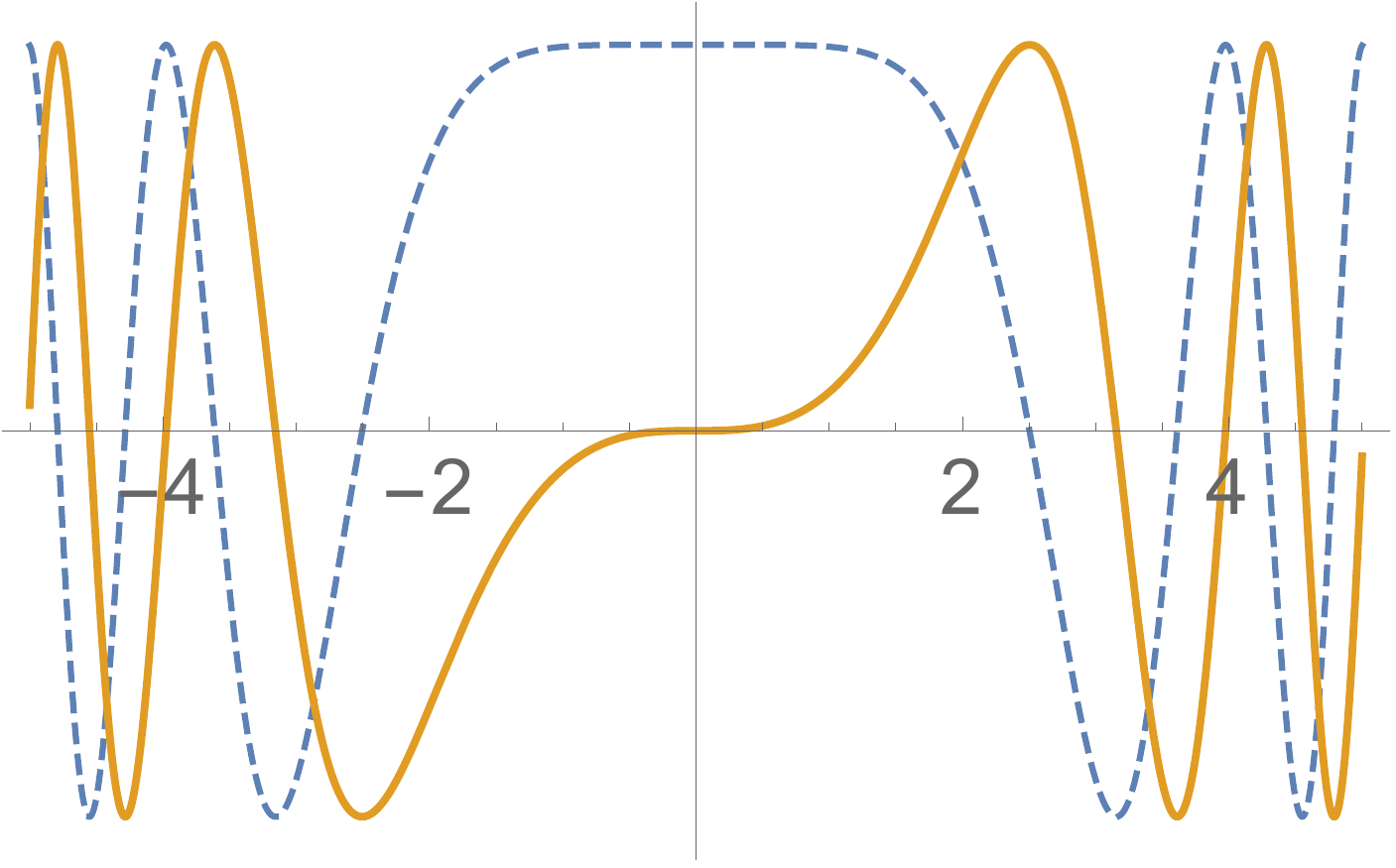}
        \label{subfig:target}
    }%
    ~ 
    \subfloat[]{
        \centering
        \includegraphics[width=0.24\textwidth]{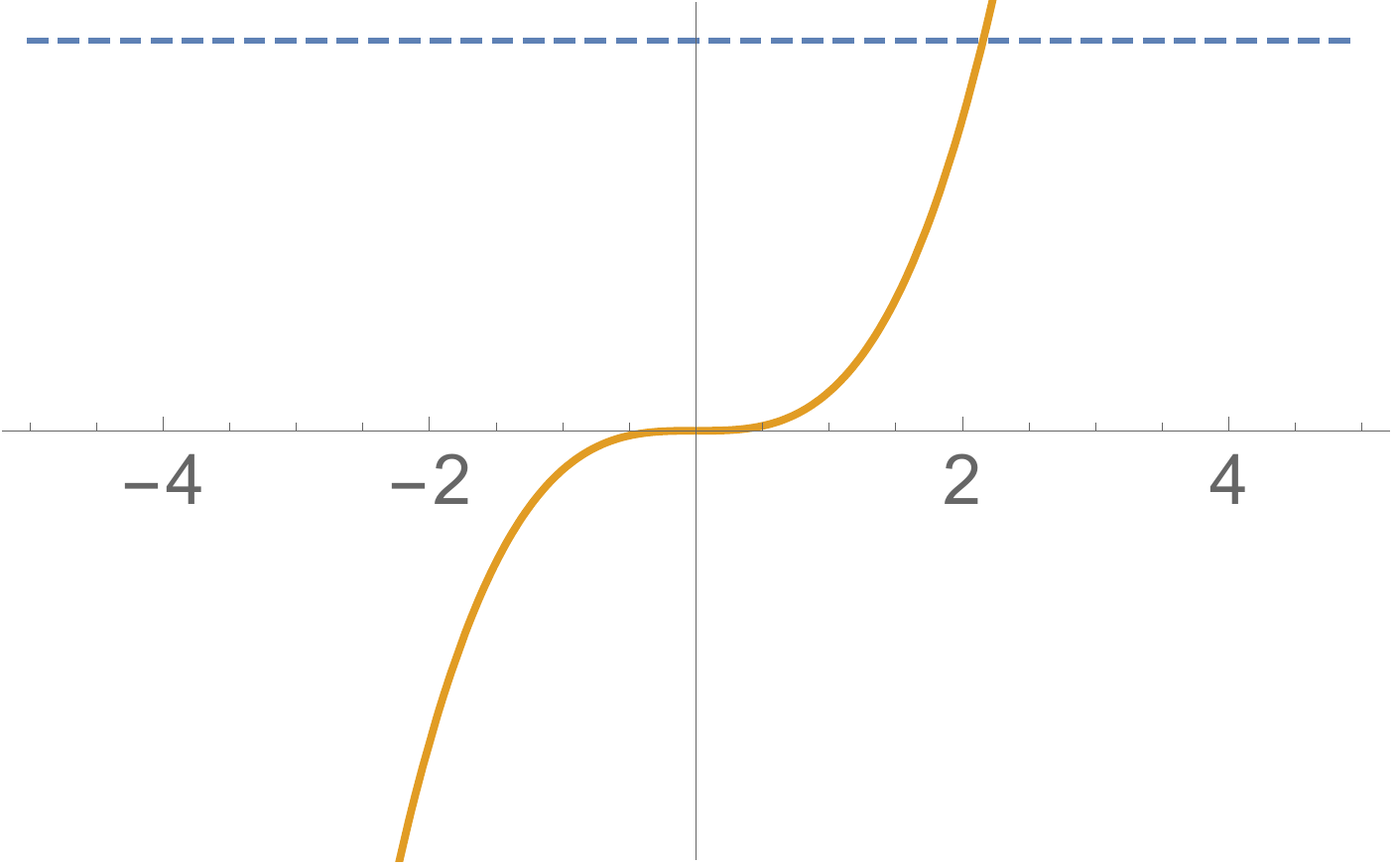}
        \label{subfig:poly}
    }
    \caption{(Color online) Position representation of the target (a) cubic phase gate and (b) its third-order expansion for $\nu=0.1$. The blue dashed lines correspond to the real parts, the yellow solid lines correspond to the imaginary parts.}
    
\label{fig:posReprGate}

\end{figure}

\begin{figure}
    \subfloat[]{
        \centering
        \includegraphics[width=0.23\textwidth]{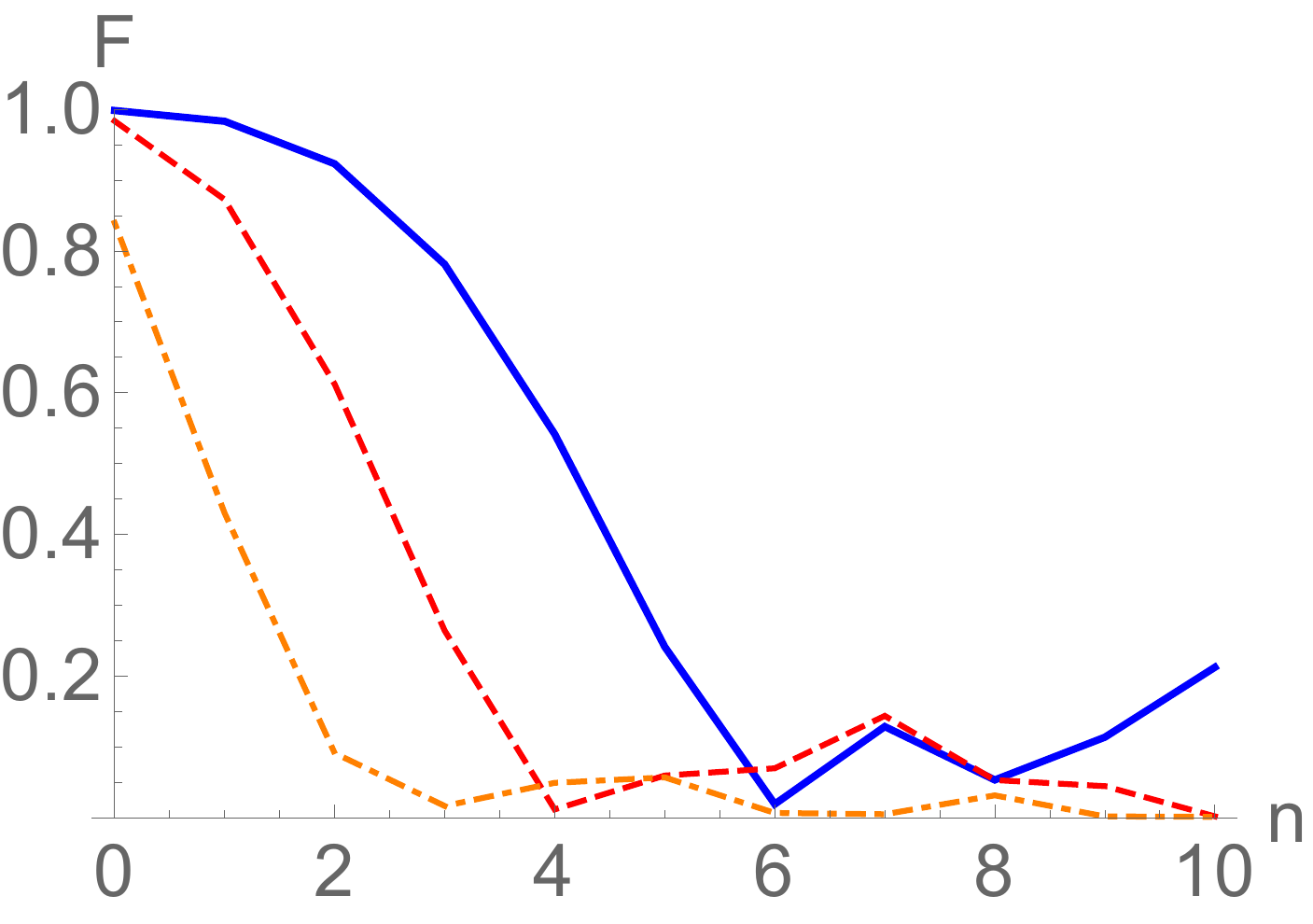}
        \label{subfig:fockClean}
    }%
    ~ 
    \subfloat[]{
        \centering
        \includegraphics[width=0.23\textwidth]{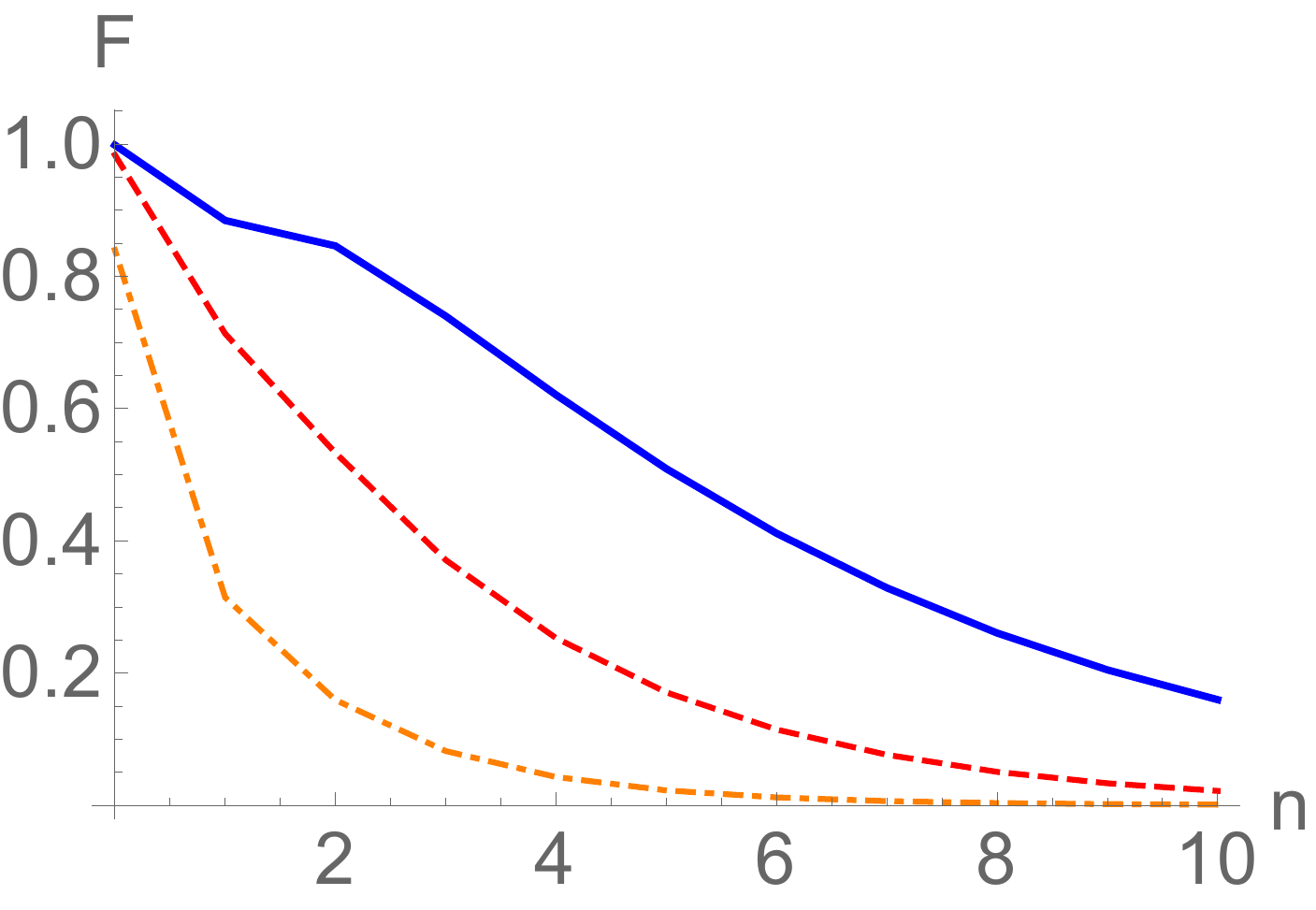}
        \label{subfig:cohClean}
    }
    \caption{(Color online) Fidelity (Eq.~(\ref{eq:fidelityPure})) between the state obtained applying either $\hat{U}=\exp\lr{i\nu\hat{q}^3}$ or its third order Taylor expansion $\hat{\mathcal{T}}$ on (a) Fock states and (b) coherent states. The $x$ axis corresponds to (a) the input photon number and (b) average photon number, respectively. The various curves correspond to three different values of the parameter $\nu$: blue solid line for $\nu = 0.1$, red dashed line for $\nu = 0.2$ and orange dot-dashed for $\nu = 0.5$.} 
\label{fig:fidClean}
\end{figure}

Being a function of the $\hat{q}$ quadrature only, the cubic phase gate is a multiplicative operator in the position representation. The real and imaginary parts of this function are plotted for the third order polynomial approximation as well as for the target unitary gate in Fig.~\ref{fig:posReprGate}, giving an indication of the quality of the polynomial approximation. Besides the imperfections of the effective gate resulting from the application of our protocols, which will be studied in the next sections, one already sees that the bare polynomial function resembles the target gate only close to the origin, so we expect it to be a good approximation only when applied to states whose position-representation wave function is concentrated around zero. Fig.~\ref{fig:fidClean} shows the fidelity of the polynomial gate with the target one for Fock and coherent states. As expected, this turns out to be better for states containing fewer photons, since their support is more concentrated around the origin. Also, the fidelity of the gate drops faster for increasing photon number when the parameter $\nu$ of the cubic phase gate is increased. This indeed  corresponds to increased evolution times, for which the Taylor expansion becomes a worse approximation. 

\section{Definitions \label{sec:BG}}

Before starting the analysis of the protocols, we recall here for reference some basic definitions and set the conventions we will use.

To each mode of the field are associated a creation operator $\hat{a}^\dagger$ and an annihilation operator $\hat{a}$, obeying the commutation relation $  \left[ \hat{a}, \hat{a}^\dagger \right] = 1$, that we use to define the quadrature operators $\hat{q} = \left(\hat{a} + \hat{a}^\dagger \right)/\sqrt{2}$ and $\hat{p} = \left(\hat{a} - \hat{a}^\dagger \right)/(i\sqrt{2})$. We will denote the quadratures' eigenstates corresponding to the eigenvalue $s$ as $\Ket[q]{s}$ and $\Ket[p]{s}$ respectively. They are related by a Fourier trasform: \begin{align} \Ket[p]{s}&=\frac{1}{\sqrt{2\pi}}\int_{-\infty}^{\infty}\ddd t e^{ist}\Ket[q]{t} = \hat{F}\Ket[q]{s} \\
\Ket[q]{s}&=\frac{1}{\sqrt{2\pi}}\int_{-\infty}^{\infty}\ddd t e^{-ist}\Ket[p]{t} = \hat{F}^\dagger\Ket[p]{s}
\end{align} which also gives $\Bra[p]{t}\Ket[q]{s}=e^{-ist}/\sqrt{2\pi} $. Any eigenstate of $\hat{q}$ can be obtained from $\Ket[q]{0}$ applying the translation operator $\hat{X}\left(s\right) = e^{-is\hat{p}}$, namely \begin{equation}
 \Ket[q]{s}=\hat{X}\left(s\right)\Ket[q]{0}
\end{equation} and similarly \begin{equation}
\Ket[p]{s}=\hat{Z}\left(s\right)\Ket[p]{0}
\end{equation} with $\hat{Z}\left(s\right) = e^{is\hat{q}}$. Introducing the displacement operator
\begin{equation}
\hat{\mathcal{D}}\left(\alpha\right) = e^{\alpha \hat{a}^\dagger - \alpha^* \hat{a}}
\end{equation} which can be expressed in terms of translation operators as \begin{equation}\hat{\mathcal{D}}\left(\alpha\right)= e^{-i\mathrm{Im}\left(\alpha\right)*\mathrm{Re}\left(\alpha\right)}\hat{Z}\left(\sqrt{2}\mathrm{Im}\left(\alpha\right)\right)\hat{X}\left(\sqrt{2}\mathrm{Re}\left(\alpha\right)\right)
\end{equation} coherent states may be expressed as \begin{equation}
\Ket{\alpha} = \hat{\mathcal{D}}\left(\alpha\right)\Ket{0}
\end{equation} where $\Ket{0}$ is the vacuum state, for which $\hat{a}\Ket{0}=0$. The vacuum state has symmetric fluctuations in $\hat{q}$ and $\hat{p}$, namely their standard deviations are $\Delta_0 ^2 \hat{q} = \Delta_0 ^2 \hat{p} = 1/2 $, saturating the Heisenberg uncertainty relations. The squeezing operator is defined as \begin{equation}
\mathcal{S}\left(k\right) = e^{-\frac{i}{2}\ln\left(\frac{k}{\sqrt{2}}\right)\left(\hat{q}\hat{p}+\hat{p}\hat{q}\right)}
\end{equation} and acts on the quadratures according to 
\be  \mathcal{S}\left(k\right)^\dagger\left( \ba{c} \hat{q} \\ \hat{p}\ea \right) \mathcal{S}\left(k\right)
=\left( \ba{cc} \frac{k}{\sqrt{2}} & 0  \\ 0 & \frac{\sqrt{2}}{k}\ea \right)  \left( \ba{c} \hat{q} \\ \hat{p}\ea \right)
\ee so that the variance of $\hat{q}$ becomes $\Delta_k ^2\hat{q} = k^2\Delta_0 ^2\hat{q} / 2$. A general squeezed state is obtained applying the squeezing and then the displacement operators to the vacuum state. We will use the notation \begin{equation} \Ket{\alpha,k} = \mathcal{D}\left(\alpha\right)\mathcal{S}\left(k\right)\Ket{0}. \end{equation} If $k\mathrm{dB}$ is the amount of squeezing in $\mathrm{dB}$, the corresponding value of $k$ is \begin{equation}
k = \sqrt{2}\times 10^\frac{k\mathrm{dB}}{20}.
\end{equation} The position-representation wave function of a squeezed state is \be\sigma_{\alpha,k}\lr{s}= \Bra[q]{s}\Ket{\alpha,k} = \mathcal{C}\exp{\left( - \frac{\left(s-q_0\right)^2}{k^2}+ip_0s \right)} \label{eq:sqWF}\ee with $\mathcal{C} = \left(k\sqrt{\frac{\pi}{2}}\right)^{-\frac{1}{2}}$, $q_0 = \sqrt{2}\mathrm{Re}\left(\alpha\right)$ and  $p_0 = \sqrt{2} \mathrm{Im}\left(\alpha\right)$.
Fock states are eigenstates of the number operator $\hat{N} = \hat{a}^\dagger \hat{a}$, so, in the optical setting, they have a  well defined photon number. Their position wave functions are given by
\be 
\label{eq:wavef-singlephoton}
\Bra[q]{s}\ket{n} = \frac{ e^{-\frac{s^2}{2}} } { \sqrt{ 2^n n! \sqrt{\pi}} } H_n\lr{s}
\ee where $H_n\lr{x}$ denotes the Hermite polynomial of degree $n$~\cite{leonhardt1997measuring}.

\section{Method 1: Photon subtracted ancilla\label{sec:photosub}}

\begin{figure}[h!]
\includegraphics[width=0.45\textwidth]{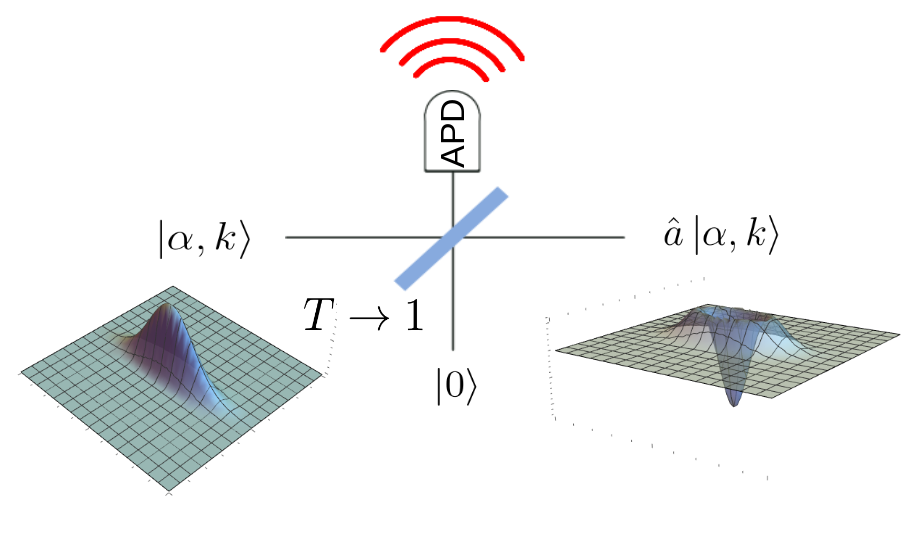}
\caption{(Color online) A method to subtract a photon from a travelling light field consists in mixing the field with vacuum in a highly transmittive beam splitter and placing a single photon detector at the output arm corresponding to the transmitted vacuum. If the transmittivity $T$ is high enough to ensure that no more than one photon is scattered from the input beam, then a click of the detector heralds a successful photon subtraction. This can be represented as the application of the annihilation operator $\hat{a}$ to the input state. The result is exact in the limit $T\to1$. }
\label{fig:expPhotoSub}
\end{figure}

For our first protocol, we exploit the idea that it is possible to induce a non-Gaussian evolution on an input state by coupling it with a non-Gaussian resource~\cite{oscill,ghose2007non, gu2009quantum, Miyata2016, KrishnaNG}. For instance, the cubic phase gate may be implemented with the help of the so-called cubic phase state \begin{equation}
\ket{\gamma\lr{\nu}} = \hat{\gamma}\lr{\nu}\Ket[p]{0}.
\end{equation} 
We focus here on photon-subtracted squeezed states as a resource. These are non-Gaussian states, displaying a negative Wigner function, whose experimental production is well established~\cite{wenger2004non,Neergaard-Nielsen2011, Ra2017}. A possible experimental implementation is depicted in Fig.~\ref{fig:expPhotoSub}. The photon subtraction can be modeled as the action of the annihilation operator $\hat{a}$.

Inspired by the basic circuit for CV-MBQC~\cite{gu2009quantum} we consider the situation described by the following circuit: 
\be \label{eq:circuit1}\Qcircuit @! {
\lstick{\ket{\psi}} & \ctrl{1} & \measureD{\hat{p}} & \rstick{m} \cw \\
\lstick{\hat{a}\Ket{\alpha,k}} & \ctrl{-1} & \rstick{ \Ket{\chi} } \qw
} \ee The input state $\Ket{\psi}$ is coupled to a photon-subtracted squeezed state $\hat{a}\Ket{\alpha,k}$ through a $\hat{C}_Z$ non demolition interaction $\hat{C}_Z = \exp\lr{ i \hat{q}_1\otimes\hat{q}_2}$ (represented by the vertical line).  The quadrature $\hat{p}$ is then measured on the first mode, giving outcome $m$. As a result, the second mode is projected on a state $\Ket{\chi}$. In the following subsection we will show that $\Ket{\chi}$ may be expressed as $\Ket{\chi} = \hat{T}_\mathrm{eff}\Ket{\psi}$ where $\hat{T}_\mathrm{eff}$ has the same form as in Eq.~(\ref{eq:genTeff}).

\subsection{Derivation of the effective transformation \label{sec:effPhotoSub}}

Neglecting for now its normalization, the output state of circuit~(\ref{eq:circuit1}) can be written as \be \Ket{\chi} \propto \Bra[p_1]{m}\hat{C}_Z\Ket[1]{\psi}\hat{a}_2\Ket[2]{\alpha,k} \ee where the projection on the eigenvector $\Ket[p_1]{m}$ of the first mode results from the homodyne measurement. Using the position representation of the operators and states involved we have  
\begin{align} \begin{split} \Ket{\chi} &\propto \Bra[p_1]{m} \hat{C}_Z\hat{a}_2 
 \int \ddd s\ddd t \psi(t) \sigma_{\alpha,k}\lr{s} \Ket[q_1]{t}\Ket[q_2]{s} \\
&\propto \Bra[p_1]{m}\hat{C}_Z  \int \ddd s\ddd t \psi(t)\left( s + \frac{\ddd}{\ddd s} \right)   \sigma_{\alpha,k}\lr{s} \Ket[q_1]{t}\Ket[q_2]{s}\\
&= \int \ddd s\ddd t e^{ist}\psi(t) f\lr{s} \frac{e^{-imt}}{\sqrt{2\pi}}\Ket[q_2]{s}
 \label{eq:firstDer} \end{split} \end{align} 
 where we made use of $\hat{a} = \frac{\hat{q}+i\hat{p}}{\sqrt{2}}$ and $\psi(t) = \Bra[q]{t} \Ket{\psi}$, with \be f\lr{s} = \left( s - \frac{2}{k^2}(s-q_0) +i p_0 \right) \sigma_{\alpha,k}\lr{s} \label{eq:fs} .\ee Recalling now that $f(\hat{q})\Ket[q]{s}=f(s)\Ket[q]{s}$ we can take the parts of the integrand depending on $s$ but not on $t$ out of the integral. The remaining integral over $\ddd s$ is the definition of the Fourier transform. We thus find \begin{align} \begin{split} \label{eq:der} 
\Ket{\chi} &\propto  f(\hat{q}) \int \ddd t \psi(t) e^{-imt} \int \ddd s \frac{e^{ist}} {\sqrt{2\pi}} \Ket[q_2]{s}\\ 
&= f(\hat{q}) \int \ddd t \psi(t) e^{-imt}  \Ket[p_2]{t} \\ 
&=  f(\hat{q}) \hat{X}\left(m\right) \hat{F} \Ket{\psi}.\end{split} \end{align} The operator $ f \lr{\hat{q}}$ can be written explitly using Eq.~(\ref{eq:sqWF}) and Eq.~(\ref{eq:fs}) as \begin{equation}
f\lr{\hat{q}} \propto \left( \hat{q} - \frac{2}{k^2}(\hat{q}-q_0) +i p_0 \right) e^{ - \frac{\left(\hat{q} -q_0\right)^2}{k^2}+ip_0\hat{q}  }. 
\end{equation} Some observations allow to simplify this expression. First, we can drop the last Fourier transform, taking $\Ket{\psi'}= \hat{F}^\dagger\Ket{\psi}$ as input. This amounts to add an inverse Fourier transform, which is just a phase-shift in the optical setting, before feeding the input to the considered circuit. We can then multiply on the left by $\mathbb{I} = \hat{X}\left(m\right)\hat{X}^\dagger\left(m\right)$ and use $\hat{X}^\dagger\left(m\right) \hat{q} \hat{X}\left(m\right) = \hat{q} + m$. Having commuted the displacement to the left, we can undo it adding a post-processing stage to our circuit consisting in a displacement depending on the homodyne outcome $m$. Finally,   
the output state $\Ket{\chi}$  has to be normalized. We introduce a normalization constant $\mathcal{N}$ depending on $k$ and $\alpha$ in which we re-absorb all numerical prefactors. As a result, the output state reads
\be \Ket{\chi} = \mathcal{N} \hat{Z}\left(p_0\right) e^{-\frac{\left(\hat{q} - q_0 +m\right)^2}{k^2} }\Big(\hat{q} - \lambda\left(\alpha,k,m \right)\Big)\Ket{\psi} \ee where 
\be 
\lambda\left(\alpha,k,m \right) =- \lr{\frac{2}{k^2 -2} } q_0 - i \lr{\frac{k^2}{k^2 -2}} p_0  - m.
\ee 
Including  a further corrective displacement in the circuit we may redefine $\hat{T}_{\mathrm{eff}}$ according to \be \label{eq:effPhotoSubCirc} \Qcircuit @C=0.9em {
\lstick{\ket{\psi}} & \gate{\hat{F}^\dagger}  & \ctrl{1} & \measureD{\hat{p}} & \control \cw  \cwx[1] \\
\lstick{\hat{a}\Ket{\alpha,k}}& \qw & \ctrl{-1} &\qw & \gate{\hat{X}^\dagger\left(m\right)} & \gate{\hat{Z}^\dagger\left(p_0\right)} &  \rstick{ \hat{T}_{\mathrm{eff}}\Ket{\psi} } \qw
} \ee 
which gives
 \be 
 \hat{T}_{\mathrm{eff}} = \mathcal{N} \exp  \left\{-\frac{\left(\hat{q} - q_0 +m\right)^2}{k^2}\right\} \Big(\hat{q} - \lambda\left(\alpha,k,m \right)\Big)  . \label{eq:effPhotoSub} 
\ee 
Note that the last correction does not depend on the outcome of the measurement and does not require adaptivity to be performed. 

The effective transformation obtained, Eq.~(\ref{eq:effPhotoSub}), is composed of two operators. The factor $\hat{q} - \lambda \lr{\alpha,k,m}$ is the desired monomial transformation. The exponential part corresponds to $\mathcal{A}\lr{\hat{q}}$ in Eq.~(\ref{eq:genTeff}). It concentrates the values of the output state wave function around the value $ q_0 - m$, which depends on the outcome of the homodyne measurement. It tends to the identity operator in the limit $k\to\infty$ corresponding to high squeezing of the ancilla in the $\hat{p}$ quadrature. However, the amount of squeezing also affects the displacements of the ancilla $q_0$ and $p_0$ that are needed to realize a target monomial for a given measurement outcome $m$. In particular, if $k \to \infty$ then it must also be $q_0 \to \infty$ if $\lambda \lr{\alpha, k, m}$ needs to have a finite real part. This has the effect to change the position of the peak of the envelope, and the resulting monomial could be distorted for high squeezing. Note that for some gates the displacements sum to zero when all the monomials in the polynomial approximation are considered. This is the case for the cubic phase gate which we study in detail. The product of the envelopes is then centered and there is no additional distortion of the gate.

It is worth mentioning that essentially the same result is found considering a photon \emph{added} rather than photon subtracted ancilla. The only difference in the above derivation consists in a minus sign before the derivative operator in the second line of Eq.~(\ref{eq:firstDer}). The effective transformation would then have the same form, just with a different $\lambda\left(\alpha,k,m \right)$. Photon addition may be easier in some experimental configurations, for example when the ancilla is only weakly squeezed, so that the average photon number is low. In that case the probability of subtracting one photon is also low.

\subsection{Gate fidelity and success probability}

As explained in Sec.~\ref{sec:poly}, one should concatenate $l$ times the circuit~(\ref{eq:effPhotoSubCirc}) to obtain an approximation of a unitary gate. The resulting transformation $\hat{ \mathcal{T} }_\mathrm{eff} \lr{\vec{m}}$ depends on the vector of the measurement outcomes $\vec{m}\in \mathbb{R}^l$ which are intrinsically random numbers. To fix the ideas, let us assume that the target polynomial is achieved for $\vec{m}=\vec{0}$. Then the effective transformation will be close to the target unitary for small values of $m_j$. The quality of the approximation as a function of $m$ can be quantified through the fidelity of the output state of the protocol with the state obtained applying the desired unitary to the input state.

Since both states are pure, we may use for the fidelity the formula in Eq.~(\ref{eq:fidelityPure}). However, the vectors $\vec{m}$ span a continuous space, hence it is not possible to post-select on a single vector, as the probability of a realization of a single vector is zero. One may consider instead an acceptance region $\Omega$ around the ideal values.  We introduce a tolerance value $\delta$ such that each stage succeeds if $|m_j|<\delta$. If at some step $|m_j|>\delta$, the protocol fails. We assume that $\delta$ is much bigger than the resolution of the homodyne detector, so that this can in turn be considered as ideal. The output state of such a procedure is hence a statistical mixture of the (normalized) states $\mathcal{T}_\mathrm{eff} \lr{\vec{m}}  \Ket{\psi}$ weighted by the probability $p\lr{\vec{m}}$ of obtaining the vector of outcomes $\vec{m}$ divided by $p_\Omega$, which is the probability of obtaining $\vec{m}$ within the acceptance region. This ensures that the output density matrix has unit trace: \be \label{eq:impureOut} \rho_\Omega = \int _\Omega \ddd ^n m \frac{p\lr{\vec{m}}}{p_\Omega}  \hat{\mathcal{T}}_\mathrm{eff} \lr{\vec{m}}  \Ket{\psi}    \Bra{\psi} \hat{\mathcal{T}}_\mathrm{eff} ^\dagger \lr{\vec{m}}. \ee The general formula Eq.~(\ref{eq:fidelityGen}) must then be used to compute the fidelity. We expect that to large values of $\delta$ correspond high success probabilities. On the other hand, large values of $m$ imply large deviations from the target polynomial, and thus a worse approximation of the desired unitary. 

\subsection{Targeting the cubic phase gate}

\begin{figure}
    \subfloat[Fock states input]{
        \centering
        \includegraphics[width=0.24\textwidth]{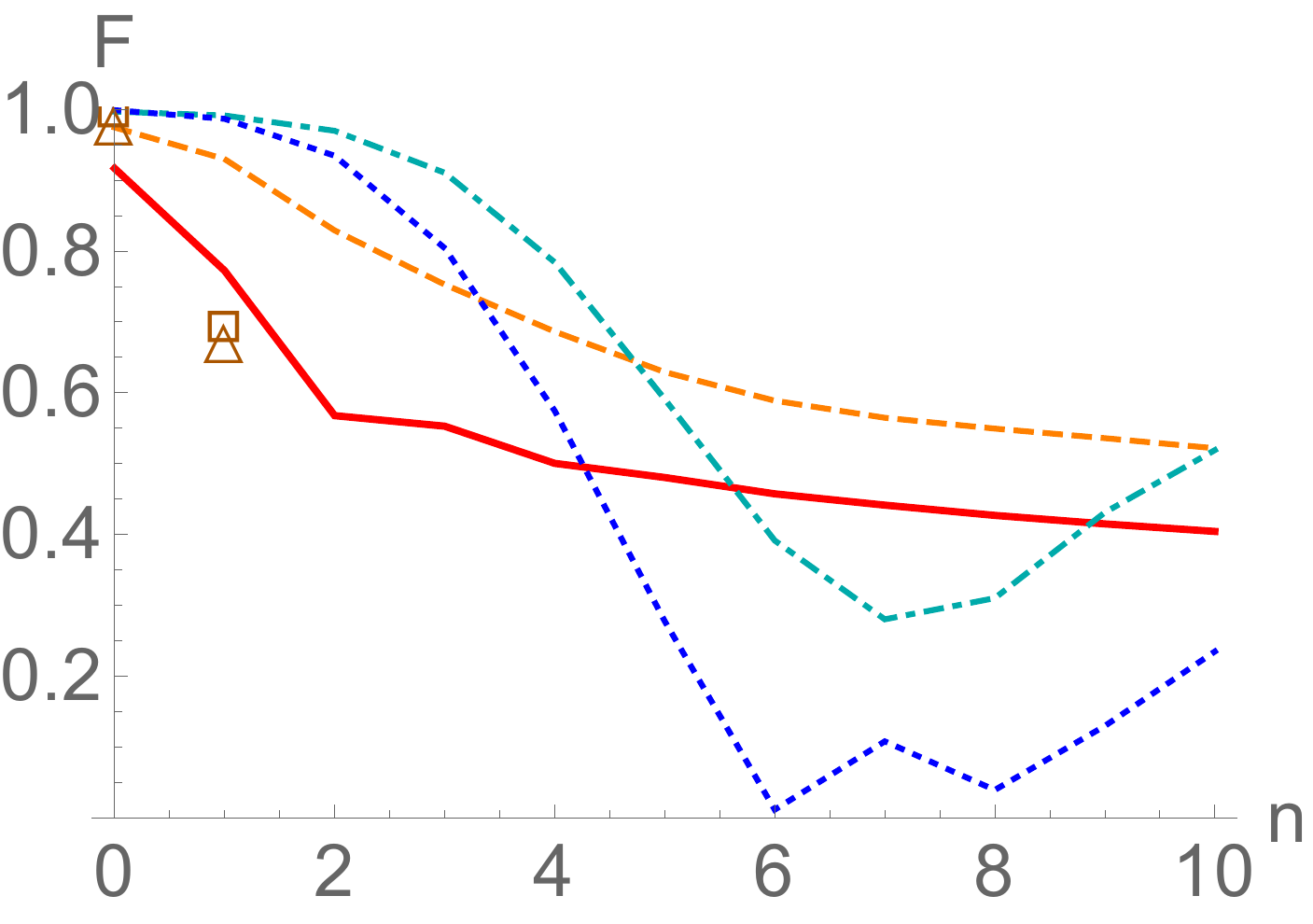}
    }%
    ~ 
    \subfloat[Coherent states]{
        \centering
        \includegraphics[width=0.24\textwidth
        ]{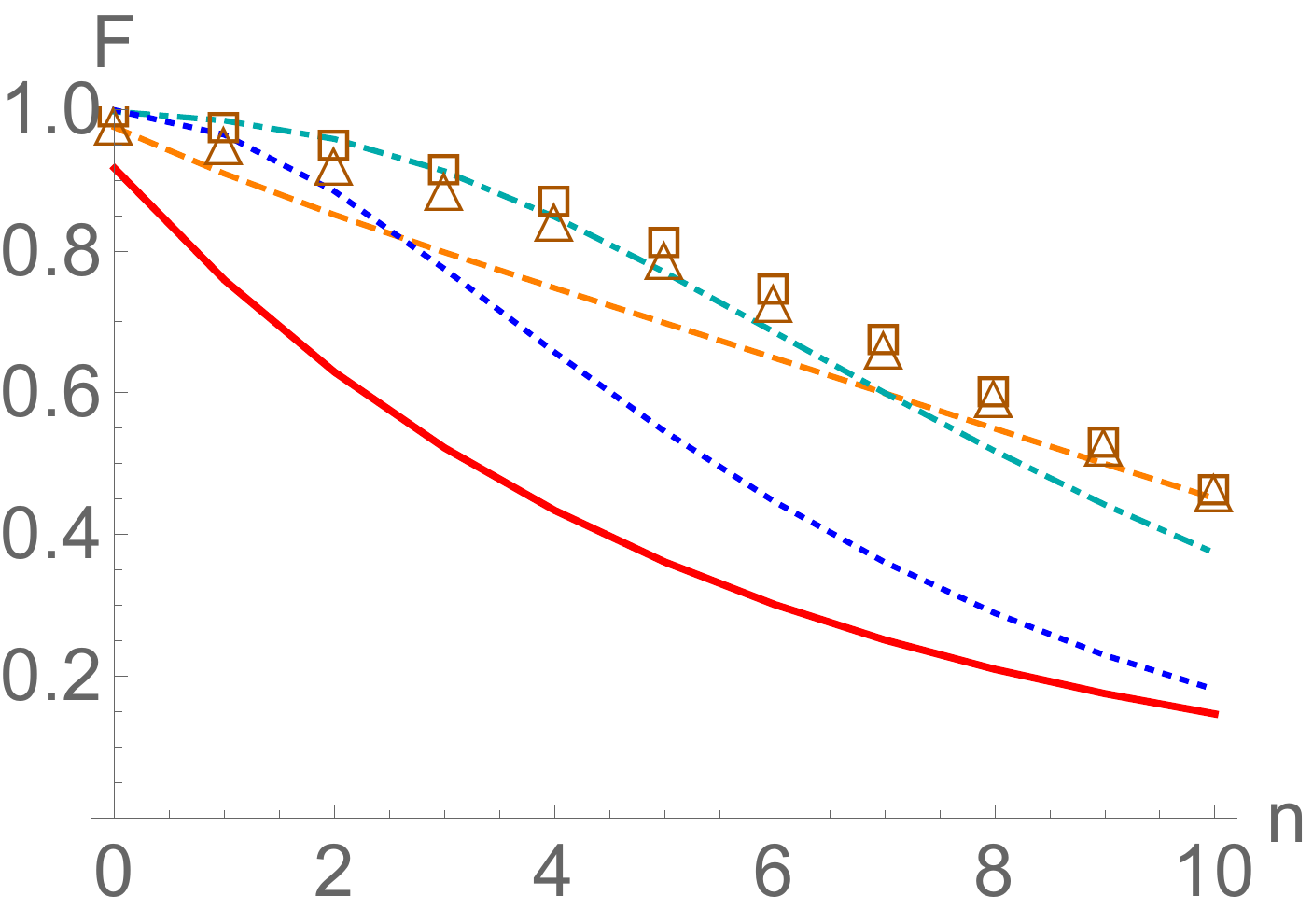}
    }
    \caption{\label{fig:pSubFid} (Color online) Method 1: fidelity between the state obtained applying the approximate cubic phase gate $e^{i 0.1 \hat{q}^3}$ and its polynomial approximation built from three sequential applications of the circuit Eq.~(\ref{eq:effPhotoSubCirc}) to Fock states (a) and coherent states (b). The lines are obtained for the specific triple of homodyne outcomes realizing the exact polynomial. The solid red line was obtained for 1 $\mathrm{dB}$ of squeezing in the ancilla, the orange dashed line for 5 $\mathrm{dB}$, the cyan dot-dashed line for 10 $\mathrm{dB}$ and the blue dotted line for 20 $\mathrm{dB}$. The orange squares represent the fidelity by post-selecting on the three homodyne outcomes in the acceptance region defined by $\delta = 0.1$ and using a 5 $\mathrm{dB}$ squeezed ancilla, while the triangles represent the same but for $\delta = 0.5$.}
\end{figure}

As anticipated, we target a cubic phase gate. For this one has to concatenate the circuit in Eq.~(\ref{eq:effPhotoSub}) at least three times. Fig.~\ref{fig:pSubFid} shows the fidelity of the approximated cubic phase gate with the ideal one for Fock states and coherent states input. In the plot, the lines represent the fidelity obtained by supposing that the perfect outcome corresponding to the desired $\lambda$ are obtained at each iteration, for various squeezing levels ranging from $1$ to $20$ dB. As it should result, the blue curve, obtained in the high squeezing limit, corresponds to the blue solid curve in Fig.~\ref{fig:fidClean}, i.e. to the fidelity of the polynomial approximation with the target cubic phase gate, because finite squeezing effect are negligeable in the implementation of our gate in this case.
However, one sees that despite being a closer approximation to the polynomial, the effective gate obtained using higher squeezing ancillae turns out not to be a better approximation of the target gate. As discussed in Sec.~\ref{sec:poly}, this is due to the fact that the polynomial itself differs from the target gate far from the origin, growing indefinitely for large $q$. This difference is attenuated faster by the Gaussian envelope if the squeezing is lower (see also discussion in Sec.~\ref{sse:discussion-envelope}).

Next, we evaluate the fidelity in the case where a finite acceptance region is considered for the outcomes of the homodyne measurement. 
Although it is possible, in principle, to compute the success probability analytically for coherent and Fock states input (see Appendix~\ref{app:fidelity1}), the calculation of the fidelity is in general computationally heavy. We then estimated it with a numerical integration method, which we could only carry out for coherent states and the single photon state. The details of the calculation can be found in Appendix~\ref{app:fidelity1}. 
We notice that for coherent states containing up to the number of photons considered here, the fidelity is higher when the post-selection occurs in a finite region, rather than on a single point. This counter-intuitive effect may be due to the complex interplay between the Gaussian envelope appearing in Eq.(\ref{eq:effPhotoSub}) and the measurement outcomes in the post-selected region.
However, as expected, the fidelity then degrades when a larger region is considered. For the single photon case the effect of post-selecting on a finite region is more detrimental.

\subsection{State preparation}

The probability of measuring all the three outcomes in the acceptance region can be very low (of the order of $10^{-9}$ or smaller in the examples considered), which makes this protocol hardly realizable in the lab. The success probability can however be improved having some \emph{a priori} information on the input state. This is due to the fact that the value of $\lambda$ in each monomial depends on the combination of the displacement in the ancilla and the homodyne outcome. Namely, from Eq.~(\ref{eq:effPhotoSubCirc}) one sees that the measurement outcome for which the correct monomial is achieved is given by \begin{equation}
m_o = - \mathrm{Re}\left[ \lambda\lr{\alpha, k, m} \right]  -\lr{\frac{ 2 }{ k^2 -2 } } q_0.
\end{equation} Since the probability of the outcomes depend on the displacement in the ancilla, one could, knowing the input state, choose the value of the displacement that maximizes the probability of getting the corresponding outcome $m_o$. This way success probabilities of the order of $10^{-4}$ can be achieved. 

\begin{figure}
\begin{minipage}{0.1\columnwidth}
        \includegraphics[width=\textwidth]{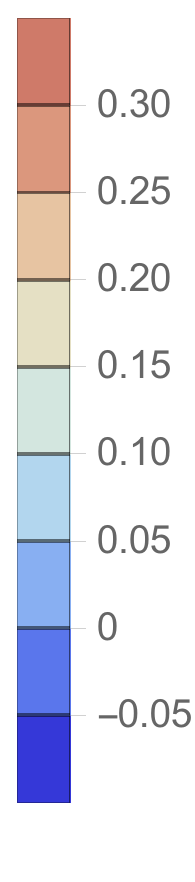} 
\end{minipage}
\begin{minipage}{0.85\columnwidth}
        \includegraphics[width=0.48\textwidth]{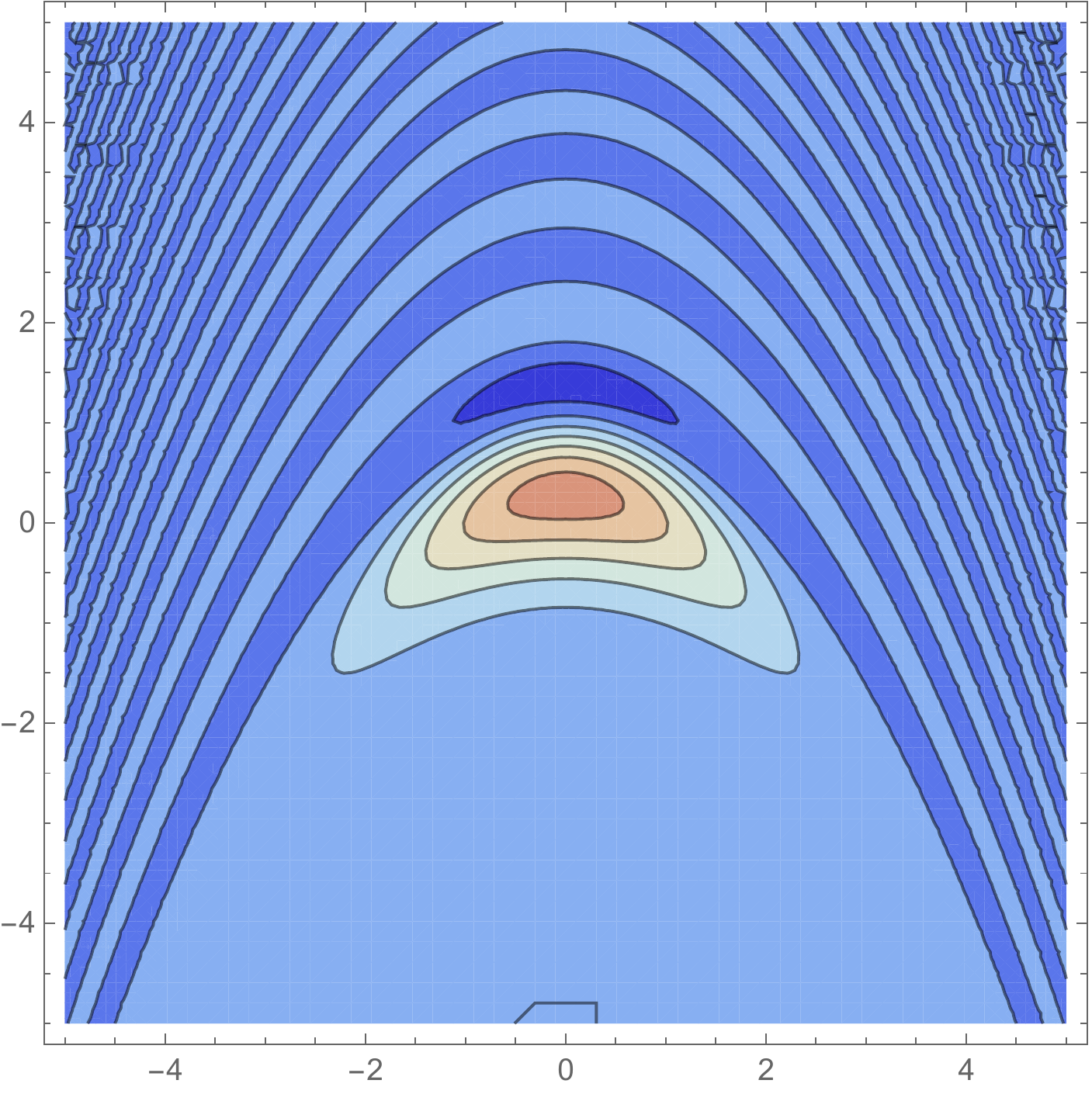}
        \includegraphics[width=0.48\textwidth]{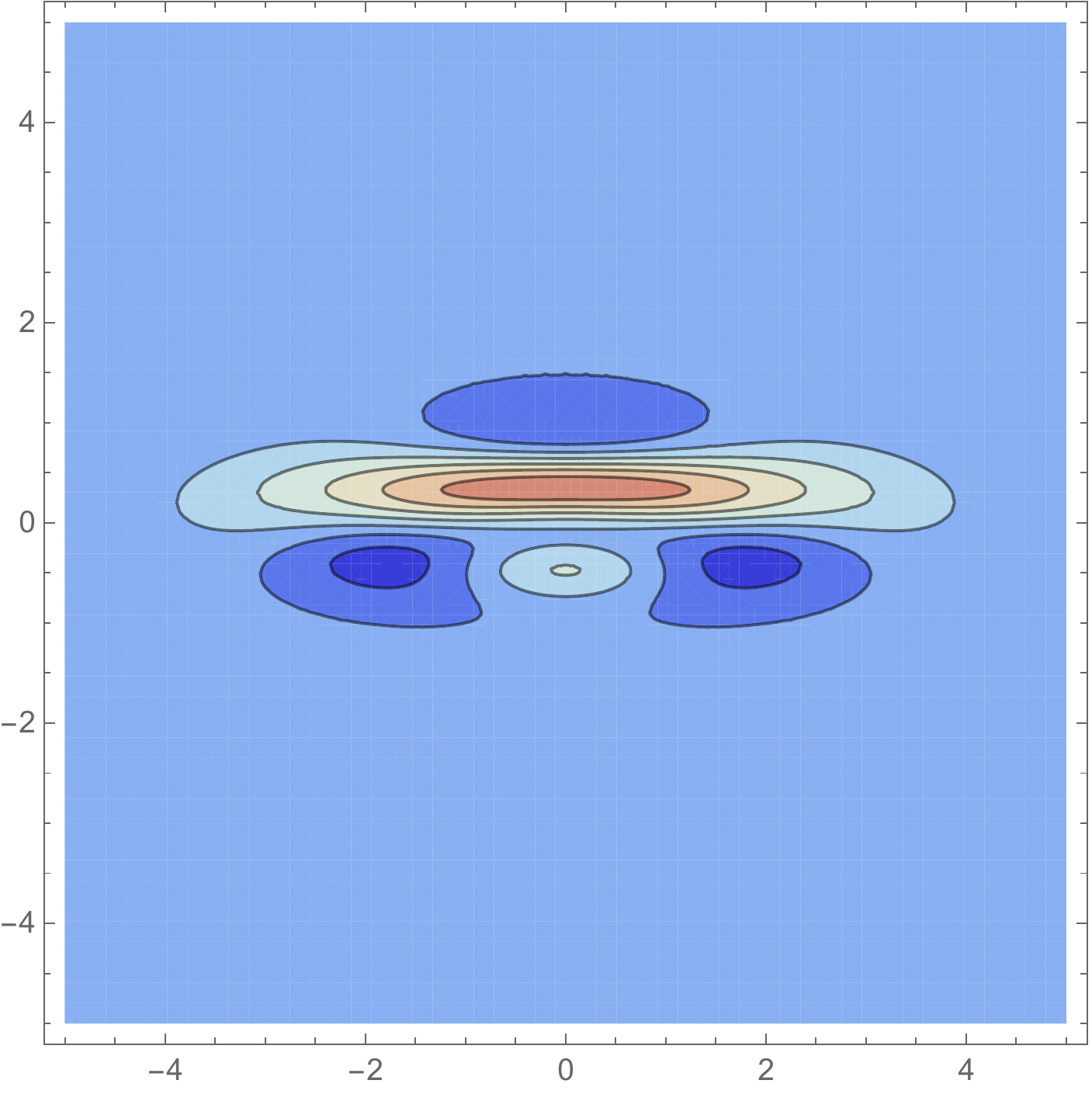} 
\end{minipage}
\begin{minipage}{0.1\columnwidth}
       \hspace{\columnwidth} 
\end{minipage}
\begin{minipage}{0.85\columnwidth}
        \includegraphics[width=0.48\textwidth]{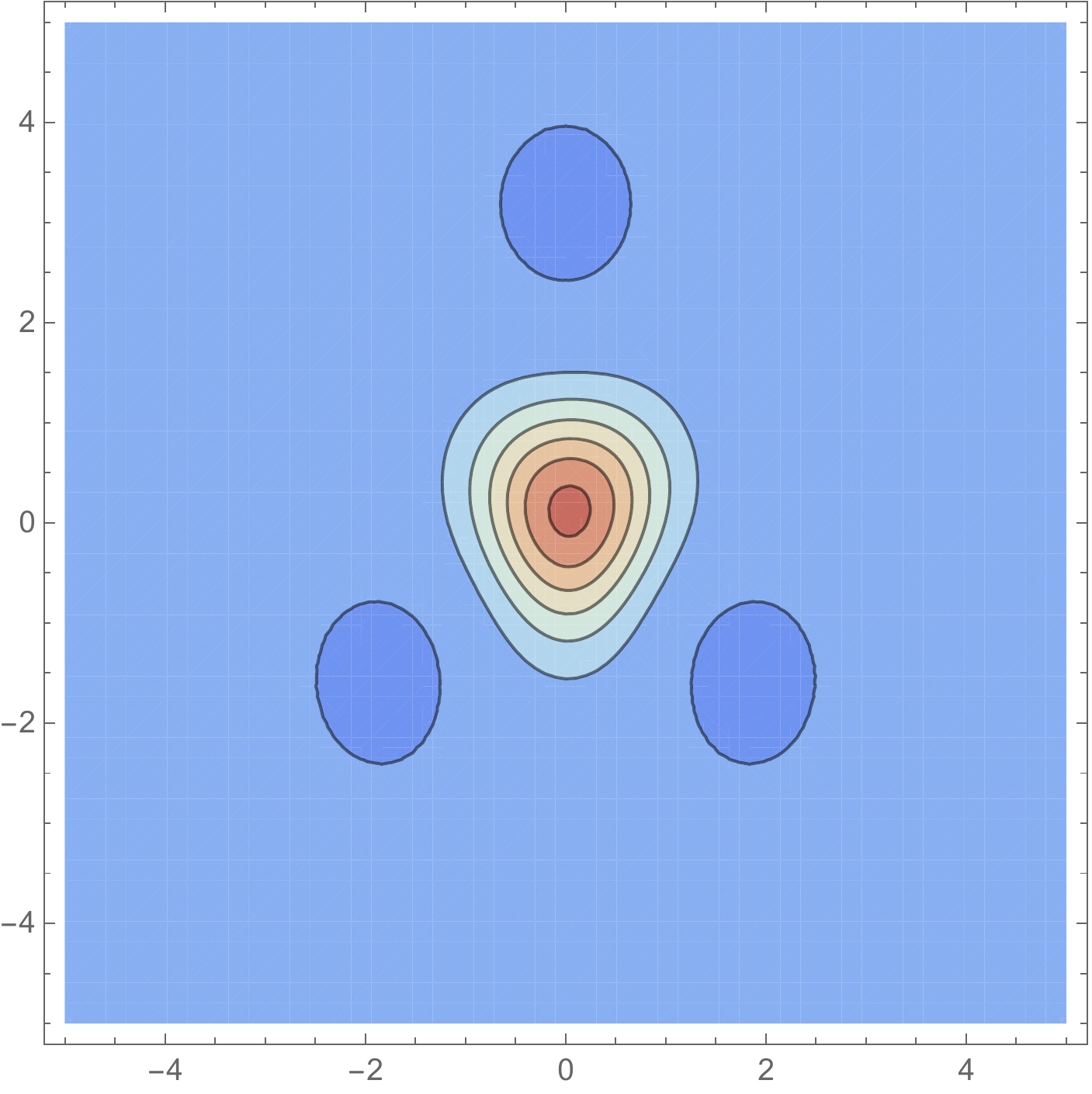}
        \includegraphics[width=0.48\textwidth]{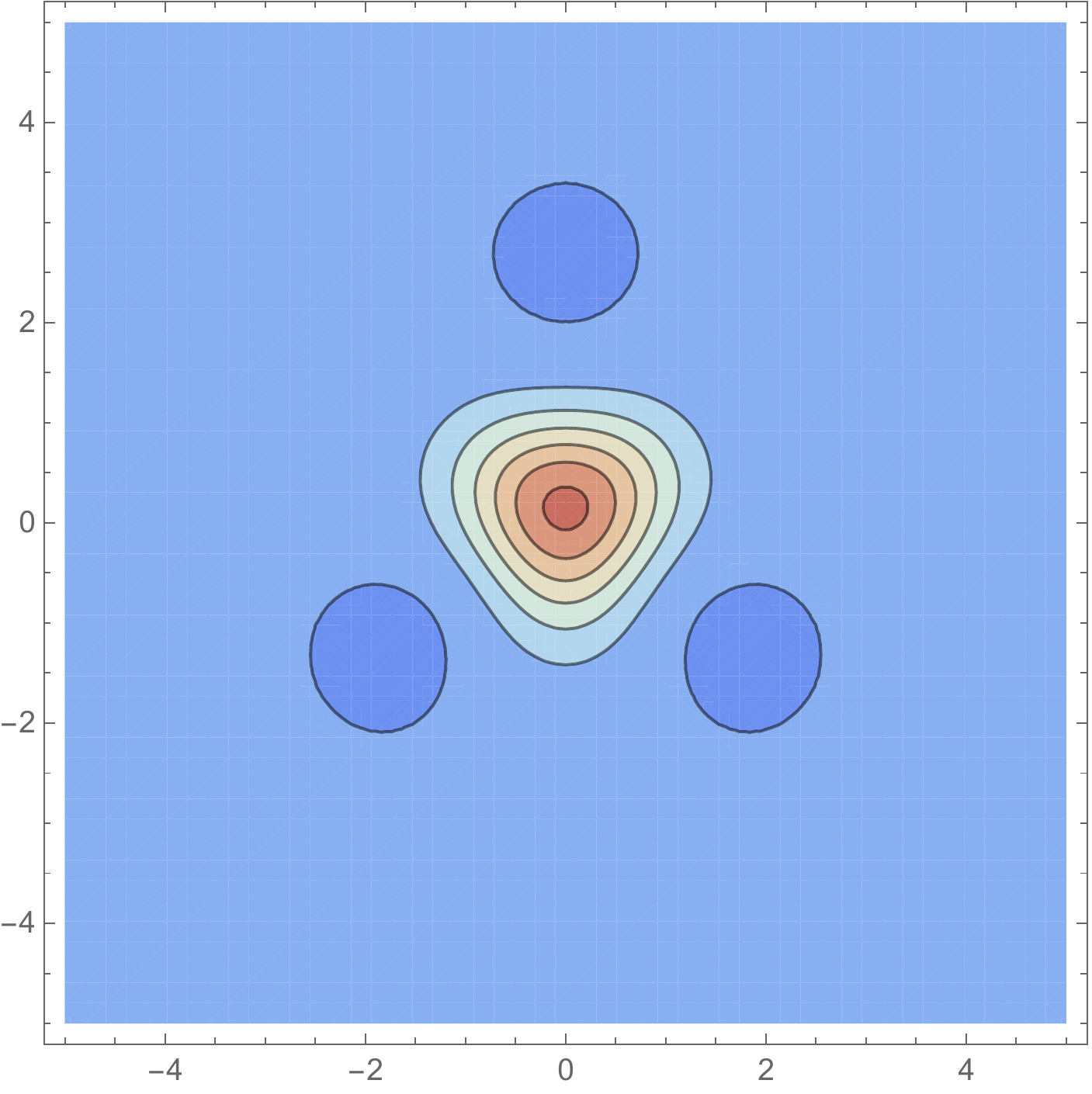}
 \end{minipage}
    \caption{\label{fig:pSubWig} (Color online) Contour plots of the Wigner functions of the states obtained applying (a) the cubic phase gate for $\nu = 0.1$ (b) its polynomial approximation (hence these two figures are independent on our protocols), (c) method 1 for exact measurement outcomes and (d) method 2, all of them for a 5 $\mathrm{dB}$  momentum-squeezed state as input state.}
\end{figure}

Ideally, any method for applying a quantum gate should be independent of the input state, but this optimized protocol can be used to improve the generation rate of a resource state, in the same spirit of what is discussed at the beginning of Sec.~\ref{sec:photosub}. For example, instead of directly applying a cubic phase gate, one could produce an approximated cubic phase state by using three sequential applications of Eq.~(\ref{eq:effPhotoSubCirc}) to an input squeezed state. 
Fig.~\ref{fig:pSubWig} shows the contour plot of the Wigner functions obtained applying the ideal cubic phase gate (a), its polynomial third order approximation (b) and our iterative protocol for exact measurement outcomes (c) to a $5$ dB squeezed state. The marked difference between (a) and (b) stems as a result of the polynomial approximation, as is also illustrated in Fig.~\ref{fig:posReprGate}. Three regions of negativity of the Wigner function obtained with the polynomial approximation are recognizable in Fig.~\ref{fig:pSubWig} (b) and are retrieved with our protocol. The fidelity of the obtained state (c) with the target state (a) is of 0.90. 

\section{Method 2: Single-photon counter\label{sec:click}}

In our second protocol, the main non-Gaussian resource is again a single-photon counter (SPC). In the previous protocol such a detector was employed to herald the production of the non-Gaussian ancilla. Here we consider instead a Gaussian ancilla, namely a squeezed state, and the SPC will replace the homodyne detector. This is represented by the circuit \be  \Qcircuit @C=3em {
\lstick{\ket{\psi}} & \ctrl{1} &  \rstick{ \Ket{\chi} } \qw \\
\lstick{\Ket{\alpha,k}} & \ctrl{-1} &\measureD{\hat{\Pi}} & \rstick{1} \label{eq:basicSPD}\cw
} \ee Note that the detection happens this time on the second mode. The positive-operator-valued measure (POVM) $\hat{\Pi}$ needs not be a full-fledged photon counter, but should be able to distinguish between no photons, one photon or more than one photon impinging on the detector. This allows to project on the one-dimensional subspace spanned by the single photon state. Such a detector is slighly more refined than a plain ``click" detector, that would only distinguish any number of photons from vacuum, causing the output state to be mixed in the case of a detection event.  The projection on the single photon state applies instead an effective transformation similar to the one derived in Sec.~\ref{sec:photosub}, as we shall see in the next section.

The output state $\Ket{\chi}$ of circuit~(\ref{eq:basicSPD}) reads \be \Ket{\chi} \propto \Bra[2]{1} \hat{C}_Z \Ket[1]{\psi} \Ket[2]{\alpha,k}.\ee The effective transformation acting on the input state $\Ket[1]{\psi}$ may be written as (neglecting normalization) $\Bra[2]{1} \hat{C}_Z\Ket[2]{\alpha,k}$ which is an operator on the Hilbert space of the first mode. This expression is the adjoint of that studied in~\cite{park}, so we expect it to induce a similar dynamics on the input state. We shall see that this is actually the case. On the other hand the physical interpretation is rather different. In Ref.~\cite{park} the effective transformation is obtained entangling a single photon with the input state and then projecting on a squeezed state. This can be done with heterodyne detection~\cite{leonhardt1997measuring}. This implies a projection on a continuous space, leading again to a trade-off between fidelity of the gate and success probability, as was the case for our first protocol. Projecting on a single photon, instead, allows for actual post-selection, since it corresponds to a well defined one-dimensional subspace, and no averaging is needed to obtain a non-zero success probability.

\subsection{Derivation of the effective transformation}

Using Eq.~(\ref{eq:wavef-singlephoton}), the output state is evaluated as \begin{align} \begin{split}  \Ket{\chi} &\propto \Bra[2]{1} \hat{C}_Z \Ket[1]{\psi} \Ket[2]{\alpha,k} \\
 &= \Bra[2]{1} \hat{C}_Z \int \ddd s \ddd t \psi\lr{t}   \sigma_{\alpha,k} \lr{s}\Ket[q_1]{t}\Ket[q_2]{s} \\
& \propto \int \ddd t \mathcal{I}\lr{t} \psi \lr{t} \Ket[q_1]{t} \end{split} \end{align} with \begin{equation}
\mathcal{I}\lr{t}  =  \int \ddd s  \sigma_{\alpha,k} \lr{s} s e^{ - \frac{s^2}{2} +ist  }  .
\end{equation} 
Evaluating the integral $\mathcal{I}\lr{t}$ we are left with a function of $t$ that can be taken out of the integral using again $\mathcal{I}\lr{\hat{q}}\Ket[q]{t}=\mathcal{I}\lr{t}\Ket[q]{t}$. We then have \begin{equation}
\Ket{\chi} \propto \hat{Z}\lr{\frac{2q_0}{2+k^2}} e^{-\lr{\frac{k^2}{4+2k^2}}  \lr{\hat{q}+p_0 }^2  } \lr{\hat{q}-\frac{2i}{k^2}q_0+p_0} \Ket{\psi} .
\end{equation} As in the case of the first protocol, we can modify the circuit~(\ref{eq:basicSPD}) adding a corrective displacement to the output state and define the effective transformation $ \hat{T}_\mathrm{eff} $  via \be \label{eq:teffSPD} \Qcircuit  @C=2.5em {
\lstick{\ket{\psi}} & \ctrl{1} &\gate{\hat{Z}^\dagger\lr{\frac{q_0}{2+k^2}}} &  \rstick{ \hat{T}_\mathrm{eff}\Ket{\psi} } \qw \\
\lstick{\Ket{\alpha,k}} & \ctrl{-1} &\measureD{\hat{\Pi}} & \rstick{1} \cw
} \ee so that it takes the form 
\be  
\label{eq:trasf-protocol2}
\hat{T}_\mathrm{eff} = \tilde{\mathcal{N}} \exp\left\{-\lr{\frac{k^2}{4+2k^2}}  \lr{\hat{q}+p_0} ^ 2 \right\} \Big(\hat{q} - \lambda\left(\alpha,k \right)\Big) 
\ee
 where $\tilde{\mathcal{N}}$ is a normalization factor depending on the input state and experimental parameters and \begin{equation}
\lambda\left(\alpha,k \right) = \frac{2i}{k^2}q_0-p_0 .
\end{equation}The effective transformation in Eq.~(\ref{eq:trasf-protocol2}) is remarkably similar to that obtained in~(\ref{eq:effPhotoSub}) for our first protocol. A first difference comes from the fact that the exponential attenuation becomes negligible in the limit $k\to0$, corresponding to infinite squeezing in the position operator. Again, the required displacement $q_0$ depends on the amount of squeezing $k$. Unlike the first method, this does not have an effect on the Gaussian envelope but it does influence the success probability, as higher values for the displacement imply a larger average photon number. At some point, this will in turn imply a smaller probability to measure exactly one photon in the second mode. The other important difference is that now, due to the absence of homodyne measurement, no random number appears in the definition of $\lambda\left(\alpha,k \right)$. This means that once a single photon impinges on the detector, the complex number in the monomial is completely determined by the experimental parameters $\alpha$ and $k$.
 
\subsection{Gate fidelity and success probability}

Even if no binning is needed, the effective transformation obtained chaining several times the process in circuit~(\ref{eq:teffSPD}) cannot match exactly the desired unitary transformation. This is due, on the one hand, to the fact that we anyway only effect a polynomial approximation of a unitary. On the other hand, each step adds a Gaussian envelope attenuating the wave function. Furthermore, detecting a single photon is by itself a probabilistic process. Therefore, a non-unit success probability is associated with the implementation of the desired transformation.

To assess the quality of the transformation we consider again the example of the cubic phase gate when the input states are either Fock states or coherent states. Specifically, for each input state $\Ket{\psi}$ we compute the fidelity (the overlap) between the state obtained applying the target unitary and the state obtained chaining circuit~(\ref{eq:teffSPD}) three times, by means of Eq.~(\ref{eq:fidelityPure}), as well as the success probability of the protocol. We assume $k$ to be fixed and compute the values of $\alpha$ such that $\lambda\lr{\alpha,k}$ matches the coefficients in the factorization of the Taylor expansion in Eq.~(\ref{eq:cubic}). 

\subsection{Targeting the cubic phase gate}
\label{sse:discussion-envelope}

\begin{figure}
    \subfloat[Fock states input]{
        \centering
        \includegraphics[width=0.24\textwidth]{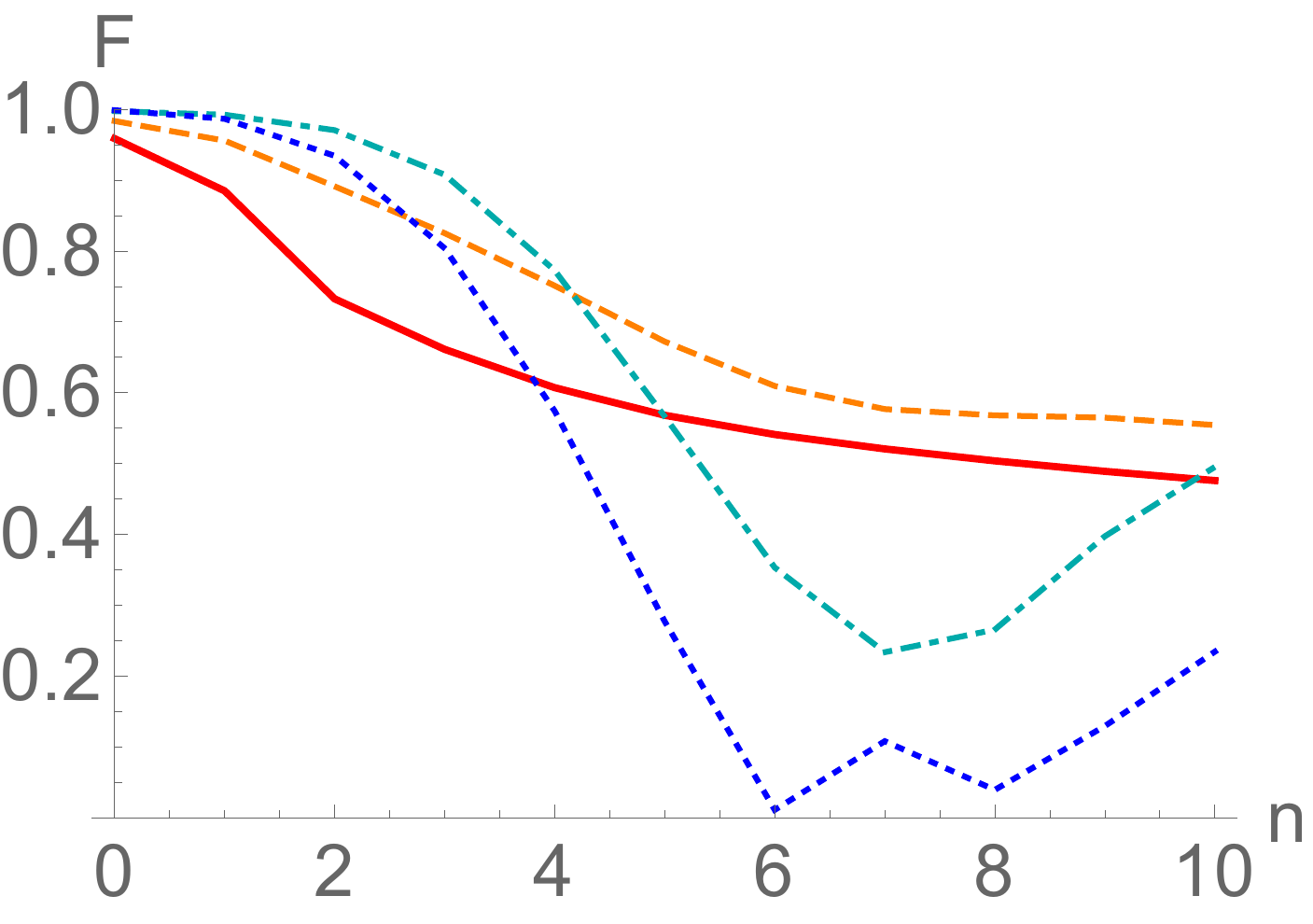}
    }%
    ~ 
    \subfloat[Coherent states]{
        \centering
        \includegraphics[width=0.24\textwidth
        ]{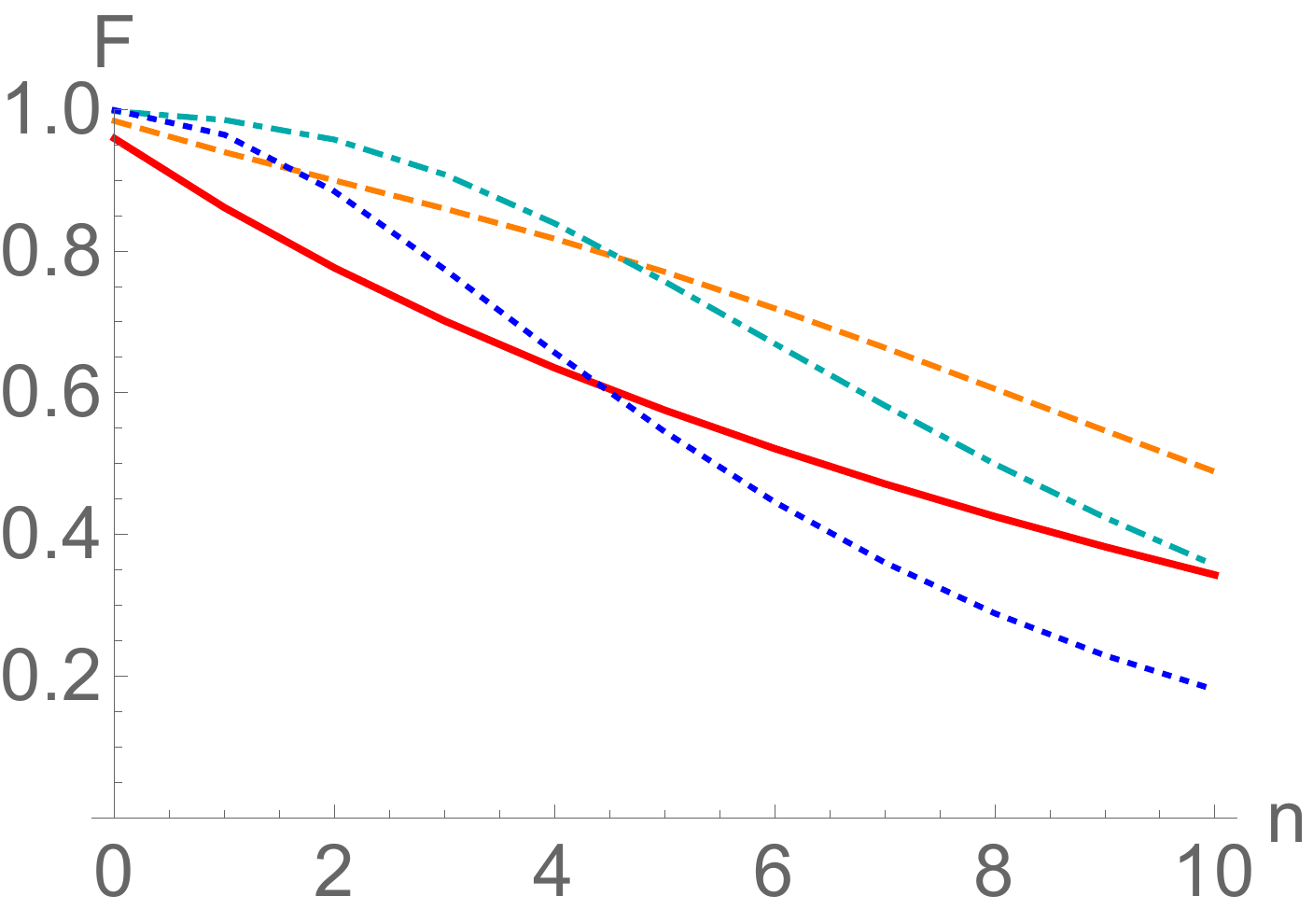}
    }
    \caption{\label{fig:SPCFid} (Color online) Method 2: fidelity between the states obtained applying either the actual cubic phase gate $e^{i 0.1 \hat{q}^3}$ or its polynomial approximation obtained with three sequential realizations of the circuit Eq.~(\ref{eq:teffSPD}) to (a) Fock states and (b) coherent states. The solid red line was obtained for 1 $\mathrm{dB}$ of squeezing in the ancilla, the orange dashed line for 5 $\mathrm{dB}$, the cyan dot-dashed line for 10 $\mathrm{dB}$ and the blue dotted line for 20 $\mathrm{dB}$. \label{fig:fidelitySPD}} 
\end{figure}

The results for the fidelity of the polynomial approximation of the cubic phase gate obtained with method two are shown in Fig.~\ref{fig:fidelitySPD}. As it was found for the first protocol, the high-squeezing case reproduces the blue solid curve of Fig.~\ref{fig:fidClean}, i.e. the fidelity of the polynomial approximation with the target cubic phase gate.  The fidelity decreases at increasing mean photon number for both Fock and coherent input states. 
As anticipated in Sec.\ref{sec:poly} (and consistently with the discussion of Fig.~\ref{fig:fidClean}), this is due to the fact that the larger the support of the input wave-function is, the more pronounced is the error intrinsic to the polynomial approximation.  This effect is sort of smoothened by the Gaussian envelope caused by finite squeezing that appears in Eq.~(\ref{eq:trasf-protocol2}): this Gaussian envelope indeed suppresses the tails of the polynomial and hence yields counter-intuitively to a better fidelity for intermediate (Fig.~\ref{fig:fidelitySPD}, orange-dashed curve) rather than high  (Fig.~\ref{fig:fidelitySPD}, blue-dotted curve) squeezing values.

\begin{figure}
    \subfloat[Fock states input]{
        \centering
        \includegraphics[width=0.24\textwidth]{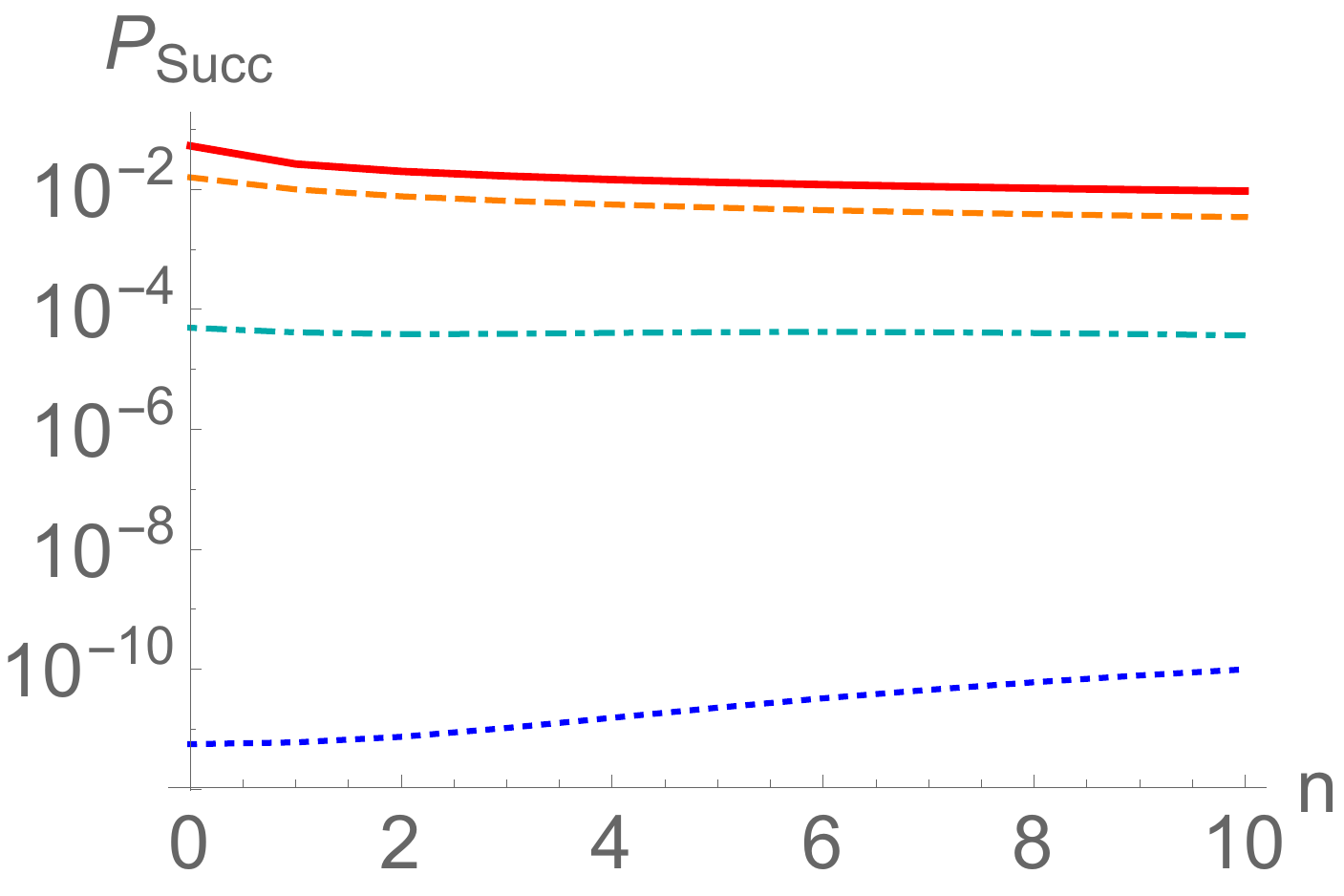}
    }%
    ~ 
    \subfloat[Coherent states]{
        \centering
        \includegraphics[width=0.24\textwidth
        ]{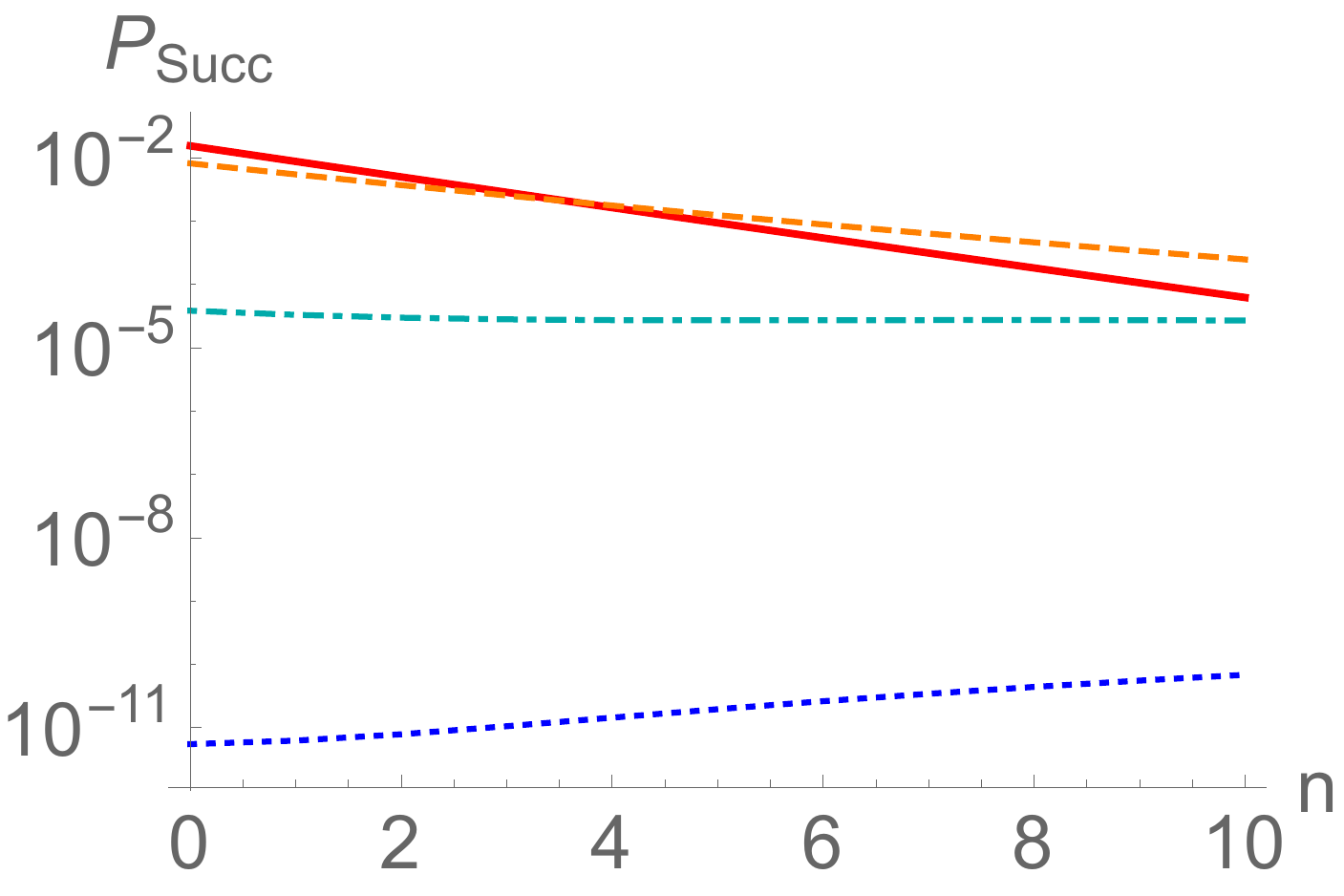}
    }
    \caption{\label{fig:succProb3} (Color online) Method 2: success probability of three sequential applications of the circuit Eq.~(\ref{eq:teffSPD}) to (a) Fock states and (b) coherent states. The solid red line was obtained for 1 $\mathrm{dB}$ of squeezing in the ancilla, the orange dashed line for 5 $\mathrm{dB}$, the cyan dot-dashed line for 10 $\mathrm{dB}$ and the blue dotted line for 20 $\mathrm{dB}$.} 
\end{figure}

The gate success probability is the product of the probabilities that a single photon is detected at each step. The details of the calculation can be found in Appendix~\ref{app:SPDProb}. The results are plotted in Fig.~\ref{fig:succProb3}. 

The probability of detecting a single photon at each iteration of the protocol is lower at larger mean photon number in the input state. As a consequence, also the success probability of the gate decreases with larger mean photon number (Fig.~\ref{fig:succProb3}). The number of photons in the ancillary squeezed states also participates to this effect: at too high squeezing, the probability of detecting a single photon at each iteration of the protocol is considerably low, so the overall success probability is also low.

We conclude that intermediate values of the squeezing in the ancillary squeezed state (between $0$ and $5$ dB for the gate we studied) are optimal for both fidelity and success probability. For these values, both fidelity and success probability are reasonably good for input states containing few photons (say up to four), and indicate that our protocol can be exploited experimentally for implementation of the cubic phase gate. 

\subsection{State preparation}

As done for the first protocol that we have presented, we target the preparation of a cubic phase state by applying the protocol outlined above to an input squeezed state. We present the obtained state in Fig.~\ref{fig:pSubWig} (d), where again we compare it to the Wigner function of the corresponding state obtained with a perfect cubic phase gate (a) as well as its polynomial approximation (b). Our protocol results in a fidelity between the retrieved state (d) and the corresponding cubic phase gate of 0.93. 

\section{Conclusions \label{sec:conclusion} }

In summary, we have presented two probabilistic protocols for engineering arbitrary evolutions diagonal in the amplitude quadrature of a single mode of the electromagnetic field, by means of a polynomial approximation. These were obtained by chaining elementary building blocks, each exploiting entanglement of the system with an ancilla and measurement.  All these operations may be achieved with existing technology.  
The spirit of our protocols is similar to that of~\cite{park}, of which they represent an alternative. Which one to choose depends on the experimental conditions. 

As an example, we refer to the experiments with frequency combs outlined in~\cite{treps} in which the relevant squeezed modes are linear combinations of frequency modes. In that case heterodyne detection of one mode would destroy the whole state, while it has been shown theoretically~\cite{Averchenko2016, Averchenko2014} as well as experimentally~\cite{Ra2017} that one or possibly more photons can be subtracted from or added to a set of squeezed modes preserving the multimode state, allowing for an implementation of our first protocol.  

However, as we have seen the typical success probabilities of this protocol are prohibitive for its actual successful implementation. Despite this fact, this protocol still retains a conceptual interest in the context of recent proposals for sub-universal models of quantum computation, such as CV Instantaneous Quantum Computing~\cite{Douce2017}. In the latter protocol, polynomial evolutions diagonal in the quadrature $\hat q$ are required as building blocks, and homodyne detections of the $\hat p$ quadrature are performed. 
Beyond the apparent match of these tools with the elements required for the implementation of our first protocol, the proof of hardness of this computational model builds on post-selection used as a mathematical trick, and therefore low success probability is not a critical issue.

The second protocol that we have presented uses single photon detection at the stage of the measurement, and results in more realistic success probabilities for a variety of input states. Therefore, it is a sensible candidate for implementations of higher-than-quadratic evolutions in the amplitude quadrature representation. Also, it could be embedded in a Measurement-Based quantum computing procedure based on the use of cluster states. This would yield an architecture where the required higher-than-quadratic order evolutions, e.g. cubic, are probabilistically implemented by means of single-photon detection on suitably chosen cluster nodes. 


\section{Acknowledgmenents}

We thank R. Filip for having shared with us the preprint of Ref.~\cite{park} prior to its publication, and P. van Loock and M. Walschaers for useful discussions.
This work is supported by the European Union Grant QCUM-bER (No. 665148), the European Research Council starting grant Frecquam and the French National Research Agency project COMB. G.F. acknowledges support from the European Union through the Marie Sklodowska–Curie Grant No. 704192. N.T. is a member of the Institut Universitaire de France.
\appendix

\section{Computation of the success probability and fidelity for a gate involving three photon-subtracted ancillae (Method 1) \label{app:fidelity1}}

We give here some details about the calculation of the fidelity between a target unitary and the polynomial gate obtained chaining three circuits of the form of Eq.~(\ref{eq:effPhotoSubCirc}). Namely, we will apply Eq.~(\ref{eq:fidelityGen}) to the state in Eq.~(\ref{eq:impureOut}). To do this we need first the probability distribution \be p\lr{\vec{m}} = p\lr{m_1,m_2,m_3}\ee of getting the outcomes $m_i$ from the homodyne measurements. The input state at the first step is $\Ket{\psi}$. At the second step the first monomial has been applied, so the input state of the second circuit is $\hat{T}\lr{\alpha_1,k}\Ket{\psi}$. Similarly, the input state of the third circuit is $ \hat{T}\lr{\alpha_2,k}\hat{T}\lr{\alpha_1,k}\Ket{\psi} $. We can thus rewrite \begin{equation} p\lr{m_1,m_2,m_3} = p\lr{m_3|m_1,m_2} p\lr{m_2|m_1}p\lr{m_1}. \end{equation} Let us denote the two-mode state right before the detection by  \begin{equation} \Ket{\Psi} = \hat{C}_Z\Ket[1]{\psi} \otimes \mathcal{M}\hat{a} \Ket[2]{\alpha,k}, \end{equation} $\mathcal{M}$ being a normalization factor for the photon subtracted state. The probability of obtaining $m_1$ at the first homodyne detection is then \begin{equation}
p\lr{m_1} = \Bra{\Psi} \lr{ \Proj[p_1]{m_1} \otimes  \mathbb{I}_2  } \Ket{\Psi}. \end{equation} This can be rewritten as \begin{equation}
p\lr{m_1} = \int\mathrm{d}x | \psi_p\lr{m_1-x} |^2\times \left| \omega_{\alpha k}\lr{x}\right|^2 \label{eq:pm1}
\end{equation}  where $ \psi_p\lr{s}$  is the wave function of the input state in momentum representation and $\omega_{\alpha k}\lr{x}=\mathcal{M}\Bra[q]{x}\hat{a}\Ket{\alpha,k}$. The expressions for the probabilities at the second and third steps are obtained replacing $\Ket{\psi}$ with $\hat{T}_1\Ket{\psi}$ and $\hat{T}_2\hat{T}_1\Ket{\psi}$ respectively. If the input state is a Gaussian pure state or a Fock state, the integrals can in principle be computed analytically. In fact, for these input states, the integrand is always of the form $\mathcal{G}\lr{x}\mathcal{Q}\lr{x}$ where $\mathcal{G}\lr{x}$ is the exponential of a second-order polynomial and $\mathcal{Q}\lr{x}$ is a polynomial. A change of variable $x = x\lr{y}$ allows to replace $\mathcal{G}\lr{x}$ with a centered Gaussian distribution $\tilde{\mathcal{G}}_\sigma\lr{y}$ of standard deviation $\sigma$, depending on $\alpha$, $k$, $m_1$ and the input state. This also maps the polynomial to \begin{equation}
\tilde{\mathcal{Q}}\lr{y} = \sum _n \gamma_n y^n,
\end{equation} with coefficients $\gamma_n$ depending on $\alpha$, $k$, $m_1$ and the input state. The integral in Eq.~(\ref{eq:pm1}) then takes the form \begin{equation}
p\lr{m_1} = \sum _n \gamma_n \mu _n
\end{equation} where $\mu _n$ is the $n$th moment of $\tilde{\mathcal{G}}_\sigma\lr{y}$  \begin{equation}
\mu _n = \begin{cases} 0 & \mbox{if } n\mbox{ is even} \\ \sigma ^n \lr{n-1}!!  & \mbox{if } n\mbox{ is odd} \end{cases}.
\end{equation} This is clearly still true for the second and third stage of the protocol, in which case the input state is  $\hat{T}_1\Ket{\psi}$ and $\hat{T}_2\hat{T}_1\Ket{\psi}$ respectively. 

Plugging Eq.~(\ref{eq:impureOut}) into Eq.~(\ref{eq:fidelityGen}), we see that the square of the fidelity of the output state obtained post-selecting on homodyne outcomes within the acceptance region is the average of the square of the fidelity for the single outcomes weighted with the respective probability \begin{equation}
\mathcal{F}_\Omega ^2 = \int \limits _\Omega  \mathrm{d} ^3 m  \frac{p\lr{\vec{m}}}{p_\Omega}  \mathcal{F}\left(\vec{m}\right)^2 \ee with \be \mathcal{F}\lr{\vec{m}} = \left| \Bra{\psi} e^{-i\nu\hat{q}^3} \mathcal{T}_\mathrm{eff}\left(\vec{m}\right) \Ket{\psi} \right|.
\end{equation}

\section{Computation of the success probability for a gate involving three single-photon detections \label{app:SPDProb} (Method 2)}

We first focus on one realization of the circuit Eq.~(\ref{eq:teffSPD}). The two-mode state right before the detection is now \be\Ket{\Psi} = \hat{C}_Z\Ket[1]{\psi} \otimes \Ket[2]{\alpha,k}. \ee 
The probability of detecting $n$ photons is given by \be p\lr{n} = \Bra{\Psi} \lr{ \mathbb{I}\otimes \Proj[2]{n} } \Ket{\Psi} \ee with $\Proj[2]{n}$ the projector on the $n$-photons Fock state of the second mode. Using Eq.~(\ref{eq:sqWF}), one gets with a few lines of algebra \be \label{eq:intApp2}  p\lr{n} = \int \limits _{-\infty}^{\infty}\ddd x \left| \psi\lr{x} \right|^2 \times \left| w_{n,\alpha,k}\lr{x} \right|^2 \ee where \be w_{n,\alpha,k}\lr{x} = \bra{n}e^{ix\hat{q}}\Ket{\alpha,k} = \int \limits _{-\infty}^{\infty}\ddd x \sigma_{\alpha,k}\lr{x}\phi_n\lr{x}e^{ixy}  \ee having denoted by $\phi_n\lr{x}=\Bra[q]{x}\Ket{n}$ the position wave function of the $n$-photon Fock state. 
The probability $p\lr{1}$ of detecting a single photon at the first step is obviously computed with the initial state as input state and setting $\alpha=\alpha_1$. At the second step, the input state is $\hat{T}_\mathrm{eff}\lr{\alpha_1,k}\Ket{\psi}$, assumed to be normalized. The function $w_{1,\alpha_2,k}$ is now used in Eq.~(\ref{eq:intApp2}). Similarly, the probability of a single photon detection at the third step has to be computed by taking as input the normalized state obtained applying $\hat{T}_\mathrm{eff}\lr{\alpha_2,k}\hat{T}_\mathrm{eff}\lr{\alpha_1,k}$ to $\Ket{\psi}$. The success probability of the three-steps protocol is given by the product of these three numbers.


\bibliography{./biblio}{}

\end{document}